\documentclass[12pt]{article}

\usepackage[capposition=top]{floatrow}
\usepackage{adjustbox}
\usepackage{setspace,graphicx,epstopdf,amsmath,amsfonts,amssymb,amsthm,versionPO}
\usepackage{marginnote,datetime,subfigure,rotating,fancyvrb}

\usepackage{enumitem}
\setlist[itemize]{noitemsep, topsep=1pt}
\setlist[enumerate]{noitemsep, topsep=1pt}

\usepackage[hyperfootnotes=false]{hyperref}
\usepackage[longnamesfirst]{natbib}
\usepackage{relsize}
\usepackage{lipsum}
\usepackage[toc,page]{appendix}
\usepackage{tikz}
\usetikzlibrary{positioning}
\tikzset{main node/.style={circle,fill=blue!20,draw,minimum size=1cm,inner sep=0pt},
            }

\makeatletter
\renewcommand{\paragraph}{%
  \@startsection{paragraph}{4}%
  {\z@}{1.25ex \@plus 1ex \@minus .2ex}{-1em}%
  {\normalfont\normalsize\bfseries}%
}

\let\oldparagraph=\paragraph
\renewcommand\paragraph[1]{\oldparagraph{#1.}}

\newcommand\blfootnote[1]{%
  \begingroup
  \renewcommand\thefootnote{}\footnote{#1}%
  \addtocounter{footnote}{-1}%
  \endgroup
}
\usdate

\usepackage{sgamevar, tikz} 
\usetikzlibrary{graphs,graphs.standard}

\usepackage{multirow,array}

\excludeversion{notes}		
\includeversion{links}          

\iflinks{}{\hypersetup{draft=true}}

\ifnotes{%
\usepackage[margin=1in,paperwidth=10in,right=2.5in]{geometry}%
\usepackage[textwidth=1.4in,shadow,colorinlistoftodos]{todonotes}%
}{%
\usepackage[margin=1in]{geometry}%
\usepackage[disable]{todonotes}%
}

\makeatletter\let\chapter\@undefined\makeatother 

\usepackage{indentfirst} 
\usepackage{jfe}          
\usepackage{bbm}

\newtheorem{thm}{Theorem}

\newtheorem{prop}{Proposition}

\newtheorem{lem}{Lemma}
\newtheorem{defn}{Definition}
\newtheorem{cor}{Corollary}

\newtheorem{example}{Example}

\begin{document}

\setlist{noitemsep}  
\onehalfspacing      

\title{Substitutability in Favor Exchange}
\author{
O\u{g}uzhan \c{C}elebi\footnote{Department of Economics, Stanford University, 579 Jane Stanford Way, Stanford, CA 94305. Email: ocelebi@stanford.edu.}
}

\linespread{1.2}


\maketitle
\thispagestyle{empty}
\vspace{-.2in}
\begin{abstract}  \noindent 
I introduce a favor exchange model where favors are substitutable and study bilateral enforcement of cooperation. Without substitutability, the value of a relationship does not depend on the rest of the network, and in equilibrium there is either no cooperation or universal cooperation. When favors are substitutable, each additional relationship is less valuable than the previous, and intermediate levels of cooperation are observed. I extend the model to allow for transfers, heterogeneous players, and multilateral enforcement. My results can explain the stratification of social networks in post-Soviet states and the adoption of different enforcement mechanisms by different groups of medieval traders.

\end{abstract}

\blfootnote{I am grateful to Daron Acemoglu, Daniel Barron, Roberto Corrao, Glenn Ellison, Joel Flynn, Drew Fudenberg, Robert Gibbons, Ali Kakhbod, Giacomo Lanzani, Juan Ortner, Parag Pathak, Heikki Rantakari, Karthik Sastry, Alp Simsek, Birger Wernerfelt, Alex Wolitzky and participants in the MIT Theory and Organizational Economics Lunches for helpful comments and suggestions. First Draft: Nov 2021.}


\clearpage

\maketitle

\setcounter{page}{1}
\section{Introduction}\label{intro}
\setlength{\abovedisplayskip}{5pt}
\setlength{\belowdisplayskip}{5pt}



In many settings, ranging from villages in rural India to trade agreements in medieval Europe, individuals and firms depend on informal mechanisms to sustain cooperation. In these settings, writing down and enforcing explicit contracts is either hard or impossible due to high costs or the lack of infrastructure, and cooperative behavior enforced by informal mechanisms is important. The enforcement of cooperation has been studied extensively, usually concentrating on how coalitions of individuals can punish those who behave in uncooperative ways and the role that the network plays in enforcing cooperation.\footnote{See \cite{wolitzky2021survey} for a recent survey of this literature, as well as other important early contributions from other disciplines such as \cite{granovetter1985economic} and \cite{coleman1994foundations}.}

This paper focuses on another role of the network; the role it plays in determining how and when players interact with each other. To fix ideas, consider three individuals, $Alice$, $Bob$, and $Carol$. Suppose that there are three different types of favors denoted by $a$, $b$, and $c$. In each period, each player is able to provide a given type of favor with some probability. In each period, a player requires a favor (of random type) with some probability and can only receive that favor from players who are able to provide it in that period. First, consider \textit{monopolistic} cooperation where favor $a$ can only be provided by $Alice$, favor $b$ can only be provided by $Bob$, and favor $c$ can only be provided by $Carol$. In this case, $Alice$ and $Bob$ interact when $Alice$ requires favor $b$ and $Bob$ is able to perform it in that period, or vice versa. As a result, the probability that $Alice$ and $Bob$ interact in any given period does not depend on whether they have a relationship with $Carol$ or not. Moreover, if cooperation between $Alice$ and $Bob$ ends for some reason, $Alice$ cannot obtain favor $b$ from $Carol$ even if they have a cooperative relationship. Therefore, if $Alice$ loses her relationship with $Bob$, she loses any benefit she can obtain from that favor type. The value that she receives from her relationship with $Bob$ does not depend on how many other players $Alice$ can interact with; it is independent of the rest of her and $Bob$'s networks.

Conversely, consider the case of \textit{substitutable} cooperation where all players can provide all types of favors with the same probability, and the player who provides the favor is randomly selected if more than one player is able to. Then the probability that $Alice$ and $Bob$ interact in any period depends on whether they are linked with $Carol$ or not.\footnote{In particular, if players need a favor with probability $\alpha$ and perform a favor with probability $p$, in monopolistic cooperation $Alice$ asks a favor type $b$ with probability $\alpha/3$ and $Bob$ can provide it with probability $p$. When cooperation is substitutable, $Alice$ needs a favor with probability $\alpha$, and $Bob$ will provide it if (i) $Bob$ is the only other player who can provide the favor in that period, which happens with probability $p (1-p)$ or (ii) both $Bob$ and $Carol$ can provide the favor, but $Bob$ ends up providing it, which happens with probability $p^2/2$.} Moreover, if the cooperation between $Alice$ and $Bob$ breaks down for some reason, and if they have a cooperative relationship with $Carol$, they can rely on her to obtain favors and losing their relationship is less important. Therefore, the value of their relationship is endogenous and depends on whether they have an ongoing favor exchange relationship with $Carol$ or not.

\paragraph{Substitutability of Favors and Bilateral Enforcement} I consider a model with finitely many players and finitely many types of favors. Substitutability of favors leads to two mechanisms that cause the network structure to determine the value each player derives from their relationships and are central to the results of this paper. First, when favors are substitutable, an additional relationship affects the provision of a given favor when it is pivotal, that is, that player is the only player who can provide that favor in that period. The probability that any given partner is pivotal, and hence the (present, discounted) value of a relationship decreases in the number of relationships and is eventually dominated by the cost of providing the favor. Second, players prefer to have relationships with those who have a high number of relationships as they require favors less frequently, and players with a low number of relationships are undesirable partners. In applications, I show how these mechanisms create stratification in networks when the players are heterogeneous, may exacerbate the inequality in the society, and explain why community (multilateral) enforcement crowds out bilaterally enforced relationships. In contrast, when favors are monopolistic (see Footnote \ref{monopolyexamples} for examples), the value of each relationship is independent from the rest of the network. 
Therefore, considering substitutability is not only more realistic in many contexts but also introduces important changes in how the network structure affects the values individuals attach to their relationships, which is at the core of the analysis.\footnote{It is reasonable to think that many favors can be performed by different individuals. For example, if an individual needs a neighbor to watch their children for a short period of time, a worker needs to change the time for their shift for a day or a firm needs to partner with another to execute a transaction, possibly there are many different individuals or firms who can help them in these situations.} 
To isolate the effect of the network and substitutability have on the values of relationships and to abstract away from the role that the network plays on the enforcement of cooperation, I first study bilateral equilibria in which cooperation is enforced by bilateral punishments.\footnote{In bilateral equilibrium, the action of player $i$ when playing with player $j$ can only depend on the previous interactions between $i$ and $j$, but not on the previous interactions with others. Bilateral mechanisms play an important role in enforcing cooperation in many settings. For example, due to the lack of markets, favor exchange (\textit{blat}) relationships were crucial for obtaining necessary goods and services in the Soviet Union and were mostly enforced through bilateral reciprocity (\cite{ledeneva1998russia}, see also the discussion in Section \ref{blat:background}). In the medieval era, traders from different cultural backgrounds formed different enforcement mechanisms. Mahgribi traders relied on a multilateral enforcement mechanism, while Genoese traders relied on a legal system and a bilateral reputation mechanism (\cite{greif1994cultural}, see also the discussion in Section \ref{maghribigenoese:background}). Finally, the large literature in relational contracts studies long-term relationships between modern firms that are mainly enforced through bilateral incentives as well as the legal system (\cite{malcomson2010relational,hendley2003mechanisms}, see also the discussion in Section \ref{sec:relational}).}

\paragraph{Stable Networks} Analyzing a game in which players can exchange costly favors over time, I characterize the \textit{stable} networks; the networks that can be sustained in bilateral equilibria where all favors are performed and all links are kept on the equilibrium path.  When favors are substitutable, as the marginal value of an additional relationship is decreasing, there is an upper bound $B^*$ such that no player can have more than $B^*$ relationships in any stable network, while the network where all players have $B^*$ relationships is stable and efficient within stable networks (Theorem \ref{T1}). In contrast, when cooperation is monopolistic, there is a sharp and unrealistic discontinuity in the predicted levels of cooperation. In particular, there is a cutoff for the discount factor such that if players are patient enough, then the complete network is stable; otherwise, the empty network is the unique stable network (Proposition \ref{monopolistic coop}). As in many settings we  observe neither full cooperation nor no cooperation, under bilateral enforcement, substitutable favors make more empirically valid and reasonable predictions compared to the previously studied monopolistic case.\footnote{For example, in the favor exchange networks in the Soviet Union  described in Section \ref{blat:background}, most individuals have a small set of others on whom they rely to obtain goods and services. Additionally, in favor exchange relationship data \cite{banerjee2013diffusion} collects  in 72 Indian villages (with an average population above 900), most individuals have at least one such relationship, while more than 95 \% of individuals have fewer than $5$. This is also observed when players are firms: \cite{malcomson2010relational} describes how some firms cultivate long-term relationships with a subset of their suppliers.}

\paragraph{Strong Stability and Transfers}Even though stability characterizes networks that can be sustained when reached, it does not take a position on which relationships will be formed.\footnote{For example, the empty network is stable since there are no favors to ask or relationships to remove.}  I study a refinement of stability, \textit{strong stability}, that requires any mutually profitable relationship to be formed, and in Theorem \ref{T2}, derive an upper bound on the number of individuals who do not have $B^*$ relationships in any strongly stable network. Moreover, this bound does not depend on the size of the society, and therefore the fraction of players who do not attain the cooperation bound goes to zero as society becomes larger. 
I also study strong stability with transfers, a further refinement of strong stability where players can make transfers, and show that transfers facilitate cooperation.

\paragraph{Heterogeneous Players and Favor Exchange Networks in Post-Soviet States} In Section \ref{postsovietsec}, I first describe the favor exchange networks in the Soviet Union with a particular focus on the transformation these networks have experienced following the increased inequality caused by the break-up of the union. This inequality has caused a shrinkage of the networks of the poor, led to a stratified network structure between the rich and the poor, and allowed the use of money to supplement connections by the rich. Motivated by this, I extend the model to allow for heterogeneous players and consider a setting with rich and poor players, where rich players can provide favors easily and are able to use transfers. I then show that my model predicts three empirically observed phenomena: (i) shrinkage of the networks of the poor, (ii) the growth of the networks of the rich to the efficient levels, and (iii) the break-up of the relationships between the two groups, and emphasize the crucial role substitutability plays in these predictions. I also illustrate how the stratification of favor exchange networks caused by the break-up of the links between the rich and the poor can exacerbate the inequality in the society.

\paragraph{Community and Legal Enforcement} Section \ref{communityenforcementsection} extends the model to incorporate community (multilateral) enforcement. I first describe the enforcement mechanisms used by two different groups of traders in the Medieval Era. Maghribi traders with collectivistic beliefs invested in sharing information and relied on multilateral punishment strategies, while Genoese traders with individualistic beliefs relied on a legal system that complements a bilateral punishment mechanism. I show that community enforcement crowds out bilateral enforcement; whenever community enforcement increases cooperation, it prevents the establishment of bilateral relationships with those outside of the community. This prediction is a direct consequence of substitutability and is in line with the lack of business association between Maghribi traders and others described in \cite{greif1989reputation}. Next, I compare three different enforcement mechanisms: community enforcement, where players invest in sharing information; pure bilateral enforcement, where cooperation is enforced only by bilateral punishments; and legal enforcement, where players invest in a legal system that complements bilateral punishments. I characterize when each of them is optimal, shedding light on the adoption of different enforcement mechanisms by Genoese and Maghribi traders to sustain cooperation \citep{greif1994cultural}.

\paragraph{Related Literature}
The literature on social cooperation builds on the contributions of \cite{kandori1992social} and \cite{ellison1994cooperation} and focuses on community enforcement. The model in this paper is based on the one in \cite{jackson2012social}, who analyze the favor exchange game and focus on renegotiation-proof networks that are ``robust to social contagion'', which means that any breakdown in cooperation between two players spreads only to their mutual neighbors. They show that these two features imply that the network is a social quilt, which is a union of cliques. \cite{ali2013enforcing} focus on the other role of networks, information propagation, and show that efficient networks are cliques and cooperation is sustained through social contagion. An important feature of their analysis is the following: Assuming that any player has at most $d$ links, cliques of $d$ are efficient.

\cite{bloch2019efficient} study a model in which agents search for partners by asking for favors from others and whenever a favor is provided, two agents form an exclusive long-term partnership and leave the network. Even though the games studied and the focus on bilateral enforcement are common between this paper and \cite{bloch2019efficient}, the possibility and importance of multiple relationships allow me to focus on the size networks that can sustain cooperation, while their focus is the characterization of efficient exclusive relationships. \cite{bendor1990norms}  analyze bilateral enforcement in a model where all players play with each other every period, allowing players to deviate in all their relationships in a single period.\footnote{For example, they show that multilateral enforcement cannot improve upon bilateral enforcement when payoffs are separable and symmetric, which is not true when players are interacting with one of their partners in each period.} Several other papers study networks and their effect on social cooperation through communication (\cite{raub1990reputation},\cite{wolitzky2013cooperation}; \cite{lippert2011networks}; \cite{balmaceda2017trust}; \cite{ali2020communication}; \cite{sugaya2021communication}) or risk sharing and informal insurance (\cite{karlan2009trust}, \cite{ambrus2014consumption}, \cite{bloch2008informal}). \cite{wolitzky2021survey} presents a recent and detailed review of this literature.  

The results in this paper complement the previous literature on social cooperation, which has mainly focused on two possible uses of networks. First, a network may reflect how players communicate with each other to exchange information about how others behave (\textit{e.g.} \cite{wolitzky2013cooperation}) in order to enforce cooperation through multilateral punishments. Second, the structure of the network may play a role in sustaining cooperation through contagion strategies. Previous papers that study favor exchange (or priosoners' dilemma) models assume that players play with all players they are linked with every period and have additively separable payoffs \citep{lippert2011networks} or each link is active (in the sense that players interact in that period) with fixed probability in each period \citep{jackson2012social, ali2013enforcing}. These assumptions imply that the value players derive from a relationship is constant, does not depend on the rest of the network, and is completely lost after the relationship ends. My focus on a different role of the network, the role it plays in determining how and when players interact with each other, allows me to study some other interesting forces, such as the ability of a player to rely on the rest of their network to compensate for lost relationships.

\cite{barron2022favor} analyze a favor exchange model where agents can accumulate wealth and show that richer agents do not have incentives to participate in favor exchange and leave the community, which they call the ``too-big-for-their-boots'' effect. My analysis complements theirs by showing that favor exchange networks become stratified when agents are heterogeneous (\textit{e.g.}, rich and poor) and demonstrates inequality has an effect even when agents stay in the community.

\cite{acemoglu2020sustaining} study a model in which agents specialized in the enforcement of cooperation may exert coercive punishment. In their model, the incentives of the specialized enforcers to carry out costly punishments are central and determine whether specialized enforcement or a mix of community and specialized enforcement is optimal. I model specialized enforcement through a legal system in a nonstrategic way, where punishment can be exerted upon the deviating agents by courts. Unless the legal system is very efficient, the use of bilateral enforcement always improves outcomes. My results on the comparison of different enforcement mechanisms complement theirs by focusing on different trade-offs such as the cost of forming communication networks and population size.


This paper is also spiritually related to the large literature on relational contracts (see \cite{malcomson2010relational} for a survey). In particular, the prediction that intermediate cooperation can be sustained by bilateral enforcement when favors are substitutable is reminiscent of the insider and outsider firms in \cite{board2011relational}. In his model, faced with relationship-specific investments and hold-up problem, the principal divides agents to two groups (insiders and outsiders), while in my model, agents can sustain intermediate levels of cooperation due to the diminishing marginal value of relationships when favors are substitutable.

\section{Model}\label{model:section}

There is a finite set $N = \{1,2,...,n\}$ of players, connected on a network described by an unweighted and undirected graph $g$, represented by the set of its links. I use $ij$ to represent the link between $i$ and $j$, so $ij \in g$ indicates that $i$ and $j$ are linked in the network $g$. $g - ij$ and $g+ij$ denote the networks obtained from $g$ by deleting and adding the link $ij$, respectively. The neighbors of player $i$ are denoted by $N_i(g) = \{j | ij \in g\}$ and the degree of player $i$ in the network $g$ is the number of their neighbors, $d_i(g) = |N_i(g)|$.

Time proceeds in discrete periods indexed by $t \in \{0,1,\ldots\}$ and $g_t$ denotes the network at the beginning of period $t$. In any period, any player needs a favor with probability $\alpha$.\footnote{For simplicity, I assume at most one player requires a favor at a given period. This does not affect the results.} $F$ denotes the finite set of different favor types. Whenever a player requires a favor, the type of the required favor is randomly determined (with uniform distribution over $F$). In any period, the player $i$ is able to provide favor type $f \in F$ with probability $p_{if}$. Thus, $j$ can provide favor $f$ to $i$ if $i$ needs a favor of type $f$ and $j$ can provide favor $f$ in that period. I assume that for each $i$, $p_{if}>0$ for at least one $f \in F$. These probabilities are collected in a \textit{favor provision matrix} $M$ described in Table \ref{table:matrix}.
\begin{table}
    \caption{Favor Provision Probabilities}
    \[
M = \begin{bmatrix} 
    p_{11} & p_{12} & \dots &  p_{1|F|}\\
    p_{21} & \ddots & & \vdots\\
    \vdots & & \ddots & \vdots \\
    p_{n1} & \dots &  \dots   & p_{n|F|} 
    \end{bmatrix}
\]
    \label{table:matrix}
\end{table}
In period $t$, if $i$ requires a favor and some $j \in N_i(g_t)$ is able to provide it, then $i$ and $j$ play a normal form game $G= \langle \{i,j\}, \{S_1,S_2\},\{u_1,u_2\} \rangle$, where $i$ is player $1$ with strategy space $S_1$ and utility function $u_1(s_1,s_2)$ and $j$ is player $2$ with strategy space $S_2$ and utility function $u_2(s_1,s_2)$. If multiple players in $N_i(g)$ can provide favor (of type $f$), then $i$ plays the game with one of those players, who is determined randomly.\footnote{The assumption that $i$ and $j$ only plays if $j$ is able to provide the favor can be interpreted as whether a player can perform a favor or not is observable by both players. This assumption makes defection observable, allows the study of stable networks, and facilitates comparison with related papers (such as \cite{jackson2012social}, which this paper builds on) that do not focus on monitoring. The diminishing marginal value of relationships under substitutable favors would still be present even when a defection is not perfectly observable, but this would require studying different equilibria with punihsments on the equilibrium path instead of the stable networks.} Therefore, a society can be described by $(N,\alpha,M,G,\delta)$.

At $t=0$, players start the game with the network $g_0$, which is common knowledge.\footnote{For each player, network $g_0$ matters only through the degree of their neighbors, and the results hold as long as the players know the identity and the degree of their neighbors at $g_0$. In particular, they do not need to know with whom their neighbors are linked with and the rest of the network.} In each period $t$, the timing is as follows:
\begin{enumerate}
    \item Players decide whether to remove any links or not.
    \item The player who requires the favor (player $i$) and the type of the required favor ($f$) is randomly determined.
    \item The set of players who can provide $f$ at period $t$ is determined according to $\{p_{jf}\}_{j \neq i}$.
    \item If multiple players in $N_i(g_t)$ can provide favor $f$, one of them (player $j$) will be randomly selected (with uniform probability) to provide the favor.
    \item Players $i$ (as player 1) and $j$ (as player 2) play the game $G$.
\end{enumerate}
Players observe with whom they are playing, the action profile in every period they play, and $N_i(g_t)$ but not $g_t$. Strategies are denoted by $\sigma_i: H_{t,i} \times \{N \setminus i\} \to \Delta(\{K,R\} \times S_1 \times S_2)$, where $\sigma(h_{t,i},j)$ is a probability distribution over $i$'s decisions on her interaction with $j$ at history $h_{t,i}$; whether they keep ($K$) or remove ($R$) the link with $j$ and which (possibly mixed) action to play if they are playing with $j$ at period $t$. 
I will study the following favor exchange game ($G_f$) described in Table \ref{table:FEgame},  and its modifications.\footnote{This game corresponds to the stage game studied in \cite{jackson2012social} when $\gamma =0$.} 
\begin{table}
    \caption{Payoffs in the Favor Exchange Game}
\begin{center}
        \begin{tabular}{cc|c|c|}
      & \multicolumn{1}{c}{} & \multicolumn{1}{c}{$C$}  & \multicolumn{1}{c}{$D$} \\\cline{3-4}
      \multirow{2}*{}  & $A$ & $(v,-c)$ & $(0,-\gamma)$ \\\cline{3-4}
    \end{tabular}
\end{center}
    \label{table:FEgame}
\end{table}
In $G_f$, player 1 asks for a favor from player 2, who either cooperates and provides the favor (C), or defects and refuses to provide the favor (D). If player 2 cooperates, player 1 gets $v$, the value of the favor, and player 2 gets $-c$, the cost of providing the favor. I assume $v>c>0$, hence doing favors is costly, but it is Pareto efficient for players to exchange favors over time. If player $2$ refuses, then the favor is not provided and player $1$ gets $0$, while player $2$ gets $ - \gamma$, where $\gamma \in [0,c)$.\footnote{The assumption that $\gamma < c$ is reasonable in our settings as even in the modern countries that have a well-functioning legal system there is a large sphere for relational contracts. Until Section \ref{communityenforcementsection}, $\gamma$ is fixed. In Section \ref{communityenforcementsection}, the society chooses $\gamma$ by paying the cost $C(\gamma)$ of maintaining a legal system with expected punishment $\gamma$.} The term $-\gamma$ denotes any additional (expected) penalty that could be imposed on a deviating player, possibly by a legal system. In some settings, such as favor exchange networks between individuals, $\gamma = 0$ is more appropriate since the interaction has no legal implications, while in some other settings, such as the court system of Genoese traders, $\gamma > 0$ is more reasonable.\footnote{If players are traders or firms, then not providing a favor would correspond to cheating the other trader or delivering low-quality goods.} 

\section{Bilateral Enforcement and Subsitutability}\label{Sec3}

\subsection{Substitutable Favors}\label{substitutablesec}

I start the analysis with the substitutable case. Section \ref{monopolisticsec} studies the opposite case of monopolistic cooperation (where each agent has a monopoly over a single type of favor) and Appendix \ref{generalm} considers more general favor provision matrices and extends the main results.

\begin{defn}
The favors are substitutable if $p_{ij} = p$ for all $i,j$ and $p \in (0,1)$.
\end{defn}

$M_s(p)$ denotes a substitutable favor provision matrix where each entry is equal to $p$. When favors are substitutable, assuming all favors are performed, players have the following expected utility for a period under network $g$:
\begin{equation}\label{eq:expectedpayoff}
    u_i(g) = \alpha v \underbrace{\Big( 1- \overbrace{(1-p)^{d_i(g)}}^{\substack{\text{Probability that} \\\text{no one is available}}} \Big)}_{\substack{\text{Probability that} \\\text{favor is performed}}} - \sum_{j \in N_i(g)} \alpha c \underbrace{\Bigg(\frac{1- (1-p)^{d_j(g)}}{d_j(g)} \Bigg)}_{\substack{\text{Probability that favor} \\\text{is performed by $i$ to $j$}}}
\end{equation}

The first term represents the benefit of having $d_i(g)$ neighbors: $i$ will need a favor worth $v$ with probability $\alpha$ and there will be at least one player who can provide the favor with probability $1-(1-p)^{d_i(g)}$. The second term is the cost of having neighbors with degrees $d_j(g)$. Any $j \in N_i(g)$ receives a favor with probability $\alpha(1-(1-p)^{d_j(g)})$. As the player performing the favor is selected randomly and all players are homogeneous in their favor provision probabilities, there is a $\alpha \frac{1}{d_j(g)} \left(1-(1-p)^{d_j(g)}\right)$ chance that player $i$ will provide it, costing her $c$. Two features of $u_i$ are direct consequences of substitutability and instrumental for the analysis.
\begin{prop}\label{prop:networkeffect}
    The following statements are true:
\begin{enumerate}
    \item Suppose that $d_j(g) = d_k(g)$, $ij \not \in g$ and $ik \not \in g$. Then $u_i(g+ik) = u_i(g+ij)$ and
    $$u_i(g+ij) - u_i(g)   > u_i(g+ij+ik) - u_i(g+ij)$$
    \item Suppose that $ij \in g$ and $jk \not \in g$. Then  $u_i(g+jk) > u_i(g)$.
\end{enumerate}
\end{prop}

 The first part shows that when favors are subsitutable, the marginal value of new relationships is decreasing, This is because a relationship affects the provision of a favor when it is pivotal, \textit{i.e.} when all others are unable to provide it and a relationship is pivotal less frequently when a player has more links. This is fundamentally different (and, in most cases, more realistic) than monopolistic cooperation, where the marginal value of a new relationship is constant and does not depend on how many links the players have. The second part shows that players prefer their neighbors to have more links. When a neighbor is well-connected, the number of other players who can provide a favor to her is higher, which reduces the probability that any given player provides it. This channel will be important for the stratification of networks with heterogeneous players. Proposition \ref{prop:networkeffect} shows that when favors are substitutable, the network structure is an important determinant of the value each player derives from their relationships and relationships are more valuable for a player when they are scarce and when the partner is well connected.

\subsection{Bilateral Enforcement}

To isolate the effect of the network on the value of each relationship, and not enforcement of cooperation, I first analyze bilateral enforcement of cooperation. 
Let $h_{t,i} \in H_{t,i}$ denote the period $t$ history of player $i$ and $h_{t,ij} \in H_{t,ij}$ denote the bilateral history between players $i$ and $j$. Formally, $h_{t,ij}$ only includes the periods and action profiles of all past interactions between $i$ and $j$.

\begin{defn}\label{bilateralmeasuredef}
A strategy profile $\sigma$ is {\bf measurable with respect to bilateral histories} if $\sigma_i(h_{t,i},j) = \sigma_i(h_{t,i}',j)$ whenever $h_{t,ij} = h_{t,ij}'$ for all $i$ and $j$.
\end{defn}\label{bilateraleqldef}
If $\sigma$ is measurable with respect to bilateral histories, then the play of $i$ when playing with $j$ only depends on the past interactions between them, and not on their past interactions with other players, ruling out community enforcement. I will analyze \textit{Bilateral Equilibrium}, a refinement of Sequential Equilibrium (SE) that requires players' strategies to be measurable with respect to bilateral histories.\footnote{A sequential equilibrium is a strategy profile and belief system in which, for every player $i$ and private history $h_{t,i}$, player $i$'s continuation strategy is optimal given her belief $\mu$ about the vector of private histories $(h_{t,j})_{j \in N}$ and there exists a sequence $\sigma^k$ with $\lim_{k \to \infty}\sigma^k = \sigma$ such that for each $k$, $\mu^k$ is derived from $\sigma^k$ using Bayes rule and $\lim_{k \to \infty} \mu_k = \mu$.}
\begin{defn}\label{bilateraleqdefn}
   $(g,\sigma)$ is a {\bf bilateral equilibrium} if:
   \begin{enumerate}
       \item $\sigma$ is measurable with respect to bilateral histories.
       \item There exists a SE $(\sigma,\{\mu(h_{t,i})\}_{h_{t,i} \in H_{t,i}})$ with strategies $\sigma$, beliefs $\{\mu(h_{t,i})\}_{h_{t,i} \in H_{t,i}}$ and $g_0 = g$.
   \end{enumerate}
    \end{defn}

    If all favors are performed on the equilibrium path, the complete network maximizes favor provision. However, players have an incentive to refuse to provide favors due to the immediate costs and delayed rewards they entail. The main goal of this paper is to characterize the \textit{stable} networks where cooperation can be sustained.

 \begin{defn}\label{defn:stable}
    A bilateral equilibrium $(g,\sigma)$ is {\bf stable} if all the favors are provided and all the links are kept on the equilibrium path.
    \end{defn}
    
A network $g$ is stable if there exists $\sigma_g$ such that $(g,\sigma_g)$ is a stable bilateral equilibrium. If a network is stable, then the payoffs on the equilibrium path are uniquely determined by the network structure without any reference to the strategies.

\subsection{Stable Networks}
\begin{defn}
A relationship $ij$ is \textbf{sustainable} at $g$ if for $x \in \{i,j\}$,
\begin{equation} \label{char}
 \frac{\delta}{1-\delta}u_x(g) - \frac{\delta}{1-\delta}u_x(g - ij)  \geq c -\gamma
\end{equation}
\end{defn}

\begin{prop}\label{stabilityprop}
A network g is stable if and only if all $ij \in g$ are sustainable at $g$.
\end{prop}
Proposition \ref{stabilityprop} shows that sustainable relationships characterize stable networks. That is, there exists a bilateral equilibrium $(g,\sigma_g)$ such that all favors are provided and all links are kept on the equilibrium path if and only if Equation \ref{char} is satisfied for all relationships in $g$. This is an important simplification in the analysis of stable networks, as instead of searching for a strategy and belief pair that constitutes a stable bilateral equilibrium under a given network $g$, we only need to check whether or not inequality in Equation \ref{char} is satisfied for all relationships in $g$.

Moreover, sustainability is a very intuitive condition. A relationship is sustainable whenever the cost of providing the favor today and keeping the link $ij$ is preferred by both players to not providing the favor but losing the link $ij$. The LHS of Equation \ref{char} is the marginal (expected, discounted) value of a relationship at $g$, which is decreasing when the favors are substitutable, while the RHS of \ref{char} is the immediate cost of providing a favor (net of punishment associated with not providing the favor), which is constant. As a result, players provide favors to each other only if their degree is low enough, which limits the extent of cooperation in any stable network and results in the following theorem.\footnote{In Section \ref{generalm} in the Appendix \ref{app:hetagents}, I relax the more general substitutable favor provision matrices and derive a similar cooperation bound.}

\begin{thm}\label{T1}
For any $(N,\alpha,p,G_f,\delta)$, there is a $B(\alpha,p,G_f,\delta) \equiv B^*$ such that if there exists $j$ with $d_j(g) > B^*$ then $g$ is not stable. If $d_i(g) = B^*$ for all $i$, then $g$ is stable.
\end{thm}


I refer to $B^*$ as the cooperation bound. To gain intuition for Theorem \ref{T1}, first note that since players prefer neighbors to have more links, a player can support the maximum number of neighbors if all those neighbors have the maximum number of neighbors themselves. The following equation represents the main trade-off for a player who has $n$ neighbors who all have $n$ neighbors: 
\begin{equation}\label{tradeoff}
\underbrace{\delta \alpha v (1-p)^{n-1} p}_{\substack{\text{Benefit from being} \\\text{helped in future}}} - \underbrace{\delta \alpha c  \frac{1-(1-p)^n}{n}}_{\substack{\text{Cost of} \\\text{helping in future}}}\geq \underbrace{(1-\delta) (c - \gamma) }_{\substack{\text{(Net) Cost of} \\\text{helping today}}}
\end{equation}

\begin{figure}
    \centering
    \caption{Cooperation Bound $B^*$}
    \includegraphics[width=0.8\textwidth]{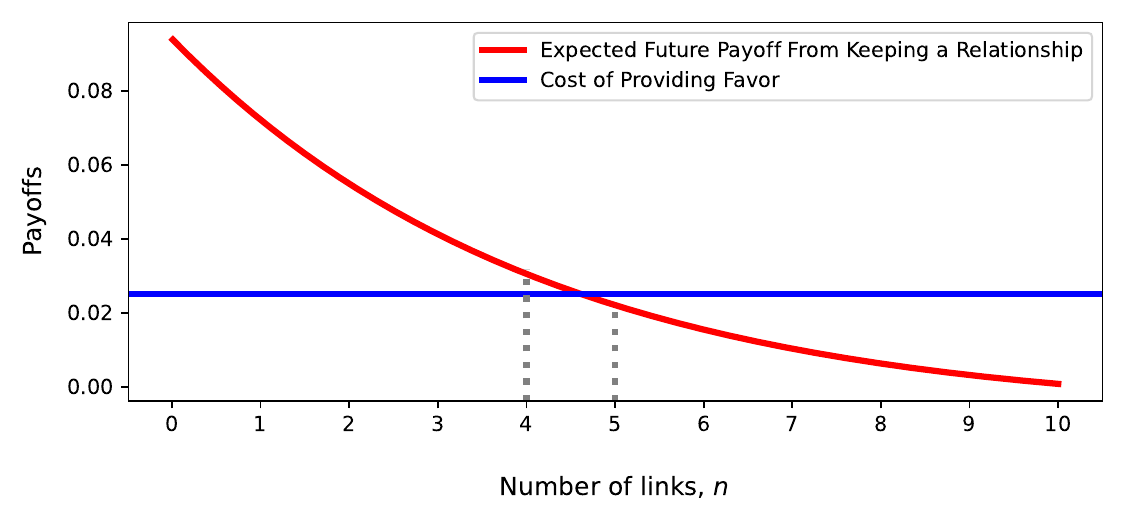}
    \label{fig:thm1fig}
    \floatfoot{\textit{Note:} Red curve plots LHS of Equation \ref{tradeoff} (present value of future payoff from a relationship), while the blue line plots the RHS (cost of providing a favor today). Curves intersect between $4$ and $5$ (dashed lines). Thus, $B^* = 4$. Parameter values: $\delta = 0.95$, $\alpha = 0.1$, $p=0.2$, $v=5.3$, $c=1.5$, $\gamma =1$.}
\end{figure}

The LHS of Equation \ref{tradeoff} is strictly decreasing in $n$ and converges to $0$, thus there exists a $B^*$ such that the inequality in equation \ref{tradeoff} holds for all $n \leq B^*$ and not for any $n > B^*$. Figure \ref{fig:thm1fig} plots the boths sides of this inequality and illustrates how $B^*$ is determined.

To see why no player can sustain more than $B^*$ links, observe that if there is a player with more than $B^*$ links in a network $g$, then the player who has the most links will not provide a favor if asked, and $g$ is not stable.\footnote{The reason for this is intuitive: let $i$ denote the player who has most links in $g$. Then $i$ would prefer not to provide a favor to any of her neighbors, since $d_i(g) > B^*$ and all neighbors of $i$ have at most $d_i(g)$ links. } Equation  \ref{tradeoff} also makes it clear that if player $i$ has fewer than $B^*$ relationships, then adding a player who has the same number of links to her network increases her payoff. As no player can have more than $B^*$ links in stable networks, the networks where all players have $B^*$ links are the most efficient stable networks. Formally, a network $g$ is \textit{constrained-efficient} if for all stable $g'$, $u_i(g) \geq u_i(g')$ for all $i \in N$. Then we have the following corollary:

\begin{cor}\label{cor1}
A network $g$ is constrained efficient if and only if $d_i(g) = B^*$ for all $i$.
\end{cor}

Although stability characterizes all possible networks that can be sustained when reached, it does not take a position on link formation. In particular, the empty network is stable as there are no relationships to end or favors to perform. One approach in the literature is to concentrate on the most efficient networks that can be sustained in equilibrium. Under that approach, if favors are substitutable, we expect to see intermediate levels of cooperation in equilibrium even in the absence of multilateral enforcement and any kind of monitoring technology.

Given the importance of the value $B^*$ for the results, it is instructive to study how it depends on the primitives of the model. As expected, cooperation becomes easier if the value of the favor is higher, the cost of the favor is lower, and the players are more patient.

\begin{prop}\label{csprop}
    $B(\alpha,p,G_f,\delta)$ is increasing in $\delta$, $v$, and $\gamma$, and decreasing in $c$.
\end{prop}

\subsection{Strongly Stable Networks}\label{strongstabsection}

This section introduces \textit{strong stability}, a refinement of stability. Strong stability requires the formation of any sustainable and mutually profitable links. When it is easy to identify and form mutually beneficial relationships, it is reasonable to expect a strongly stable network to emerge in equilibrium.
\begin{defn}\label{obtainabledef}
A network $g'$ is obtainable from g via deviations by $\{i,j\}$ if the following hold
\begin{itemize}
    \item[(i)] $kl \in g'$ and $kl \not \in g$ $\implies$ $\{k,l\}=\{i,j\}$
    \item[(ii)] $kl \in g$ and $kl \not \in g'$ $\implies$ $\{k,l\} \cap \{i,j\} \neq \emptyset$
\end{itemize}
\end{defn} 

Definition \ref{obtainabledef} characterizes all networks that $i$ and $j$ can create by forming a link between themselves and potentially removing their links with other players. Two players $i$ and $j$ \textit{violate strong stability} if they can form a mutually beneficial and sustainable link.
\begin{defn}\label{strongstability}
    Two players $i,j$ violate strong stability at $g$ via $g'$ if $g'$ is obtainable from $g$ via deviations by $\{i,j\}$, $ij \in g'$ and $ij \not \in g$ and the following hold:
    \begin{itemize}
        \item[(i)]$ij$ is sustainable at $g'$
        \item[(ii)] $u_i(g')  \geq  u_i(g)$, $u_j(g')  \geq u_j(g)$, and at least one of the inequalities is strict.
    \end{itemize}
\end{defn}
A stable network $g$ is \textit{strongly stable} if there does not exist $\{i,j\}$ that violates strong stability at $g$.\footnote{Condition (i) is necessary since two players with same number of links can always add an unsustainable link that makes both of them better off.} The following  theorem characterizes the strongly stable networks:


\begin{thm}
\label{T2}
For $B^*$ in Theorem \ref{T1}, the following statements are true:
\begin{itemize}
    \item If $d_i(g) = B^*$ for all $i$, then $g$ is strongly stable. 
    \item In any strongly stable network, for any positive integer $k < B^*$, there can be at most $k+1$ players with $k$ links.
\end{itemize}
\end{thm}

The first part shows that constrained-efficient networks are strongly stable. Therefore, they are within the predictions of the model when profitable relationships can be identified and formed. However, there may still be some players who have fewer relationships than the efficient level. These players who do not reach the bound are in some sense undesirable partners; they require favors more frequently than they provide them. The following example illustrates this mechanism:

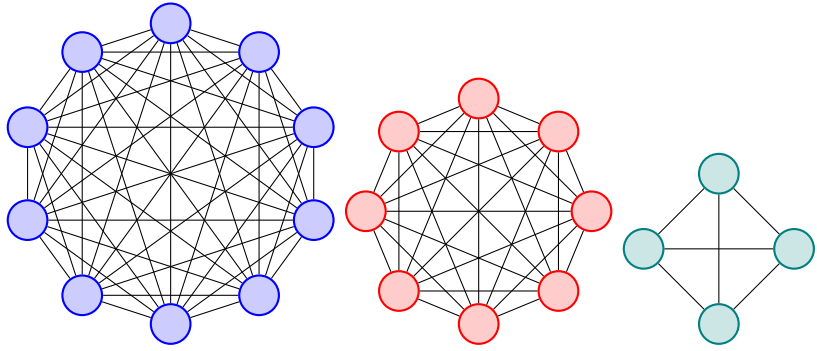
\begin{figure}
\caption{Strongly Stable Network in Example \ref{ex:ssnetwork}}
\tikzstyle{mynode}=[thick,draw=blue,fill=blue!20,circle,minimum size=15]
\tikzstyle{mynode2}=[thick,draw=red,fill=red!20,circle,minimum size=15]
\tikzstyle{mynode3}=[thick,draw=teal,fill=teal!20,circle,minimum size=15]
\tikzstyle{oldnode}=[circle,draw,fill=black]
    \centering
\begin{tikzpicture}
  \graph[empty nodes,nodes=mynode]{subgraph K_n [n=10,clockwise,radius=2cm]};
\end{tikzpicture}
\begin{tikzpicture}
  \graph[empty nodes,nodes=mynode2]{subgraph K_n [n=8,clockwise,radius=1.5cm]};
\end{tikzpicture}
\begin{tikzpicture}
  \graph[empty nodes,nodes=mynode3]{subgraph K_n [n=4,clockwise,radius=1cm]};
\end{tikzpicture}
    \label{fig1}
\end{figure}

\begin{example}\label{ex:ssnetwork}
Let $p=0.25$, $v=7$, $c=1$, $\alpha=0.15$, $\gamma = 0$ and $\delta = 0.99$. It is useful to rewrite Equation \ref{char} to denote whether a link between $i$ (with $m = d_i(g)$) and $j$ (with $n = d_j(g)$) is sustainable or not:
\begin{equation}
      h(m,n) =  -c + \frac{\alpha \delta}{1-\delta} \left(v\left(\left(1-p\right)^{m}-\left(1-p\right)^{m+1}\right)-\frac{c}{n+1}\left(1-\left(1-p\right)^{n+1}\right)\right)
\end{equation}
A link between $i$ and $j$ is sustainable at $g$ if $h(d_i(g),d_j(g)) \geq 0$. Under the given parameters, $B^* = 9$, while and $h(8,4) < 0$. Therefore, the network in Figure \ref{fig1}, where three groups of players denoted by blue, red, and green nodes form three completely connected components of different sizes, is strongly stable. In this network, each blue player has $9$ relationships. As $B^*=9$ and all blue players are linked to only other blue players, no blue player can be part of a coalition that violates strong stability. Each red player has $7$ links and each green player has $3$ links. The only possible violation of strong stability may come from a proposed link between a red and a green player. However, as $h(8,4) < 0$, any links between these two groups would not be sustainable, and the network is strongly stable. 
\end{example}

This example illustrates how networks can exacerbate inequality, even in symmetric models. Having a low degree makes players worse off not only because they are receiving favors less frequently but also because it makes it harder for them to form new, sustainable relationships. Moreover, they have relationships with others who have low degrees, which further reduces their payoffs. We will revisit this mechanism again in Section \ref{postsovietsec} and demonstrate how it causes the stratification of social networks with heterogeneous players.

The second part of the proposition limits the number of players who do not have $B^*$ links. To see the intuition behind this result,  note that if $i$ and $j$ have $n<B^*$ links, $g+ij$ is sustainable and makes both players better off. Thus, they must be linked in any strongly stable network. As $B^*$ does not depend on $N$, we obtain the following corollary, which shows that in large societies, almost all players reach the cooperation bound $B^*$ when mutually profitable relationships are formed.

\begin{cor}\label{cor2}
As $N \to \infty$, in strongly stable networks, the fraction of players who do not attain the cooperation bound vanishes.
\end{cor}
\subsection{Transfers}\label{transfersec}
In many settings, players can compensate each other in ways other than by performing the required favors. For example, if the players represent firms and favors represent a service provided by the firms, then parties can compensate each other through transfers. In such settings, the quality of the provided service may be imperfectly observable or noncontractible, so firms can still deviate by providing a subpar service, a low quality product, or refusing to share the profits, which will correspond to not providing favor.

To consider transfers, I define a further refinement of strong stability, \textit{strong stability with transfers}. Suppose that some subset of players $\tilde N \subseteq N$ have the ability to make transfers. Consider a new relationship between $i$ and $j$, where $x \in \{i,j\}$ pays $t_x$ to the other player each time they provide a favor, and $t_x > 0$ only if $x \in \tilde N$. This means that whenever $i$ and $j$ play, they play the following game:

\begin{table}
    \caption{Payoffs in the Favor Exchange Game with Transfers}
\begin{center}
        \begin{tabular}{cc|c|c|}
      & \multicolumn{1}{c}{} & \multicolumn{1}{c}{$C$}  & \multicolumn{1}{c}{$D$} \\\cline{3-4}
      \multirow{2}*{}  & $A$ & $(v-t_i,-c+t_i)$ & $(0,-\gamma)$ \\\cline{3-4}
    \end{tabular}
            \begin{tabular}{cc|c|c|}
      & \multicolumn{1}{c}{} & \multicolumn{1}{c}{$C$}  & \multicolumn{1}{c}{$D$} \\\cline{3-4}
      \multirow{2}*{}  & $A$ & $(v-t_j,-c+t_j)$ & $(0,-\gamma)$ \\\cline{3-4}
    \end{tabular}
\end{center}
     \floatfoot{\textit{Note:} When $i$ asks a favor from $j$ (left pane). When $j$ asks a favor from $i$ (right pane).}
\end{table}

Let $\tilde u_i(g,j,t_i,t_j)$ denote the expected payoff per period of player $i$ if this new relationship is added to some network $g'$, where $g = g' + ij$.\footnote{\label{footnote:transfer}To compute $\tilde u_i(g,j,t_i,t_j)$, note that the expected per period transfer $i$ pays to $j$ is $\mathbb E_g[t_i] = w(i,g) t_i$ where $w(i,g) =\frac{\alpha (1-(1-p)^{d_i(g)})}{d_i(g)}$. Then given $i$, $j$, and $t = (t_i,t_j)$ the expected payoff per period of $i$ at $g$ is $\tilde u_i(g,j,t_i,t_j) = u_i(g) - \mathbb  E_g[t_i] + \mathbb E_g[t_j]$.} First, I extend the definition of a sustainable relationship to incorporate transfers.

\begin{defn}
    The relationship $ij$ is sustainable at $g$ with transfers $t=(t_i,t_j)$ if
    \begin{equation}\label{eq:sustainabilityi}
        \frac{\delta}{1-\delta}(\tilde u_i(g,j,t_i,t_j) - u_i(g-ij)) \geq c- \gamma
    \end{equation}
        \begin{equation}\label{eq:sustainabilityj}
        \frac{\delta}{1-\delta}(\tilde u_j(g,i,t_j,t_i) - u_j(g-ij)) \geq c- \gamma
    \end{equation}
\end{defn}

Next, I extend the definition of strong stability violation.

\begin{defn}\label{strongstabilitytransfer}
    Two players $i$ and $j$ violates strong stability with transfers at $g'$ via $g$ and $t$ if $g$ is obtainable from $g'$ via deviations by $\{i,j\}$, $ij \in g$ and $ij \not \in g'$ and the following hold:
    \begin{itemize}
        \item[(i)]$ij$ is sustainable at $g$ with $t =(t_i,t_j)$
        \item[(ii)] $\tilde u_i(g,j,t_i,t_j) - u_i(g') \geq 0$, $\tilde u_j(g,i,t_j,t_i) - u_j(g') \geq 0$, with at least one inequality strict.
    \end{itemize}
\end{defn}

A stable network $g$ is \textit{strongly stable with transfers} if there are no agents $i,j$ and transfers $t$ such that $i,j$ and $t$ violate strong stability with transfers. Strong stability with transfers further refines strong stability and is equivalent to it when $\tilde N = \emptyset$. The following proposition shows that transfers facilitate cooperation and allow players who can use them to reach the cooperation bound.

\begin{prop}\label{transferprop}
Suppose that $|\tilde N| \geq B^*$. In a strongly stable network with transfers, all players in $\tilde N$ have $B^*$ links (the same value in Theorem \ref{T1}). 
\end{prop}

The reason behind this proposition can be explained as follows. When a player can make transfers, she can compensate others for the favors they provide. This allows the player to avoid situations where new relationships are difficult to form due to a lack of initial connections. When all players can make transfers, in any strongly stable network, all players attain their cooperation bound.

\begin{cor}\label{transfercor}
Suppose that $\tilde N = N$. A network is strongly stable if and only if all players have $B^*$ links. 
\end{cor}

\subsection{Comparison with the Monopolistic Benchmark}\label{monopolisticsec}

This section formally compares substitutable favors with the previously studied monopolistic counterpart. Consider the setting where the number of agents is the same as the number of favors ($|N| = |F|$) and let $M = p I \equiv M_m(p)$, where $I$ is the identity matrix and $p \in (0,1)$. Under $M_m(p)$, player $i$ can only provide favor $k$ if and only if $i = k$, in other words, each player has a monopoly over their type of favor, and I refer $M_m(p)$ as \textit{monopolistic cooperation}. Furthermore, two linked players $i$ and $j$ interact if $i$ ($j$) requires favor $j$ ($i$), which happens with probability $\alpha/N$, and $j$ ($i$) is able to provide the favor that period, which happens with probability $p$, so $i$ and $j$ interact with probability $2 \hat p$ where $\hat p = \frac{\alpha p}{N}$, which is independent of the rest of their networks. Thus, any model that determines which two linked players play in each period randomly from the uniform distribution is equivalent to monopolistic cooperation in my setting.\footnote{\label{monopolyexamples}\cite{jackson2012social} assumes that in each period, any two linked players can interact with an exogenously given probability, \cite{lippert2011networks} assumes that all players play with each other in all periods, while \cite{ali2013enforcing} assumes that each link is active with an exogenously Poisson rate. These models correspond to monopolistic cooperation in the setting of this paper.} The following result helps explain why the earlier literature concentrated on multilateral enforcement rather than bilateral enforcement in monopolistic models. 

\begin{prop}\label{monopolistic coop}
Let $\hat p = \frac{\alpha p}{N}$. Under monopolistic cooperation:
\begin{itemize}
    \item If $c - \gamma \leq \dfrac{\delta}{1-\delta} \mkern9mu \hat p (v-c)$, then complete network is stable and efficient.
    \item If $c - \gamma > \dfrac{\delta}{1-\delta} \mkern9mu  \hat p (v-c)$, then the empty network is the unique stable network.
\end{itemize}
\end{prop}

The LHS of the inequality corresponds to the immediate cost savings a player can obtain by refusing to provide the favor, while RHS corresponds to the expected present value of future cooperation with any given player. Proposition \ref{monopolistic coop} shows that in terms of networks that can be supported by bilateral enforcement, there is a sharp and unrealistic discontinuity under monopolistic cooperation: either the complete network is stable, and thus universal cooperation can be sustained in equilibrium, or the empty network is the unique stable network; hence, any cooperation is impossible. Moreover, which state prevails is determined by comparing the immediate cost of the favor with the expected present value of the relationship. As expected, cooperation is possible if players are patient, favors are more valuable and less costly, the penalty for defection is stronger, and the probability that players interact in the future is higher.


\subsection{Discussion: Relationships Between Firms}\label{sec:relational}

Before moving on to extensions and applications, I will briefly discuss how my results relate to the literature on relational contracts. First, relational contracts serve as an important motivation to focus on bilateral enforcement. Empirical work has shown that relational contracts enforced by bilateral punishments play an important role in sustaining cooperation in multiple settings. \cite{mcmillan1999dispute} interviews firms in Vietnam and report that even though relational contracts play an important role in sustaining cooperation, only 19 \% of the firms would expect third-party sanctions if they are deviated by one of their partners. \cite{hendley2003mechanisms} provides evidence for the use of bilateral enforcement mechanisms is common in a study of 254 Romanian firms. They measure the importance of different mechanisms for supporting sales agreements between firms and find that $56 \%$ of the weight was on bilateral mechanisms, while enforcement that relied on third parties accounted for less than $11 \%$. \cite{murrell2003firms} analyzes under what conditions bilateral enforcement is preferred and finds that bilateral enforcement is used when the frequency of interactions is higher and the quality of the product is important but unverifiable.

The present model is different from the standard models that study relational contracts, which usually divides players into two as a firm (principal) and suppliers (agents), \textit{e.g.}, \cite{board2011relational}. However, my results can still generate insights about those settings. In particular, Theorem \ref{T1} shows that when favors are substitutable, having too many relationships means that each relationship is used less frequently and thus harder to sustain through future punishments. A similar reasoning applies to firms: If suppliers are interchangeable, then a firm with a given size can only sustain such relationships with a bounded number of suppliers. This offers an alternative explanation for some firms choosing to conduct business with a limited number of long-run partners in addition to more pronounced reasons such as the costly formation of relationships and relation-specific investments \citep{asanuma1989manufacturer}.

\section{Heterogeneous Players and Social Networks in Post-Soviet States}\label{postsovietsec}
In this section, I first give a short historical background on the favor exchange networks in the Soviet Union, focusing on the bilateral nature of the enforcement and the transformation these networks experienced after the dissolution of the Soviet Union. Second, I extend the model to allow for two types of rich and poor players, where rich players can make transfers and incur a lower cost when they provide a favor. I show that my model predicts the stratification and polarization of the favor exchange networks that arose following inequality in the post-Soviet era. Appendix \ref{morehet} considers more general formulations of heterogeneity and obtains similar results on the stratification of networks.

\subsection{Favor Exchange (Blat) Networks in Soviet Union}\label{blat:background}

\textit{Blat} is defined as the use of personal networks and informal contacts to obtain goods and services in short supply. It is a reciprocal relationship in which people exchange favors over time and was prevalent in the Soviet Union, where perpetual conditions of shortage and lack of access to diverse goods and services through the markets necessitated the use of such relationships.\footnote{A similar institution, \textit{Guanxi}, is present in China. As in blat relationships, Guanxi is characterized by an informal and personal connection between two individuals who adhere to an implicit psychological contract to maintain a long-term relationship based on interactions following dynamic reciprocity and long-term equity principles (see \cite{chen2004intricacies} for more details on Guanxi, as well as the bilateral nature of the relationships).}  These favor exchange networks were necessary to acquire many services ranging from basic necessities and quality healthcare to leisure activities such as traveling or attending concerts.\footnote{For example, the weekly goods and services a Soviet lawyer needs to obtain by blat include ``food, dry cleaning, toilet paper, concert tickets and flowers'' \citep{Simis}.}  \cite{ledeneva1998russia}, based on 56 interviews of individuals involved in these relationships, describes this phenomenon as
\begin{quote}
Blat was oriented to different needs in different historical periods (it
was already flourishing in the 1930's) ... ``Blat was simply a necessity for a decent life. You
couldn't eat or wear what you bought in the shops, everything was in
short supply, queuing and bad quality of services were appalling. To
live normally, one had to have acquaintances and informal access to
every sphere where needs arose'' many respondents remembered.
\end{quote}
Therefore, many individuals used their position in society to be useful to others who could reciprocate.  An important aspect of these relationships is that the main enforcement mechanism was bilateral. The continuation of cooperation is primarily enforced by the threat of denial of future favors. The relationship continues as long as both parties benefit from it and is enforced by the threat of losing future exchanges.\footnote{Many interviewees who used blat emphasized this nature of the enforcement. An interviewee says: \textit{ ``The exchange is fully dependent on the interest which each side has in the 'other' and previous exchanges''}. Another interviewee claims:
    \textit{``\ldots request had to be made in such a way that it would be fulfilled or, in the case of refusal, would not jeopardize the whole
relationship''}, and another describes the possibility of punishment as: \textit{``\ldots the request is not adequate, one runs the risk of being refused or even losing the relationship''.}} These observations are summarized in \cite{ledeneva1998russia} as:
\begin{quote}
 Because blat tends to be repetitive and often operated with known partners and because of the absence of any sanctions outside the relationship, it is possible to speak of balance in blat relations.
\end{quote}
\subsection{Favor Exchange Networks in Post-Soviet Countries}\label{postsovietdesc}
The favor exchange networks have experienced changes in the market economy that followed the dissolution of the Soviet Union. For example, market reforms led to richer individuals cutting ties with poorer ones in post-Soviet Russia.\footnote{Two respondents in \cite{ledeneva1998russia} describe the effect of the increased inequality as     \textit{``People separate when there is a material barrier. All my friends are now businessmen, and they turned their backs on me. Not at once,
they gradually distanced themselves.''} and     \textit{``Those who have became wealthy have dropped out of my circle.''}. Similarly, \cite{el2009you} note how abrupt market reforms and increased poverty in Jordan caused poorer people to withdraw from social networks, contributing to social and economic polarization.} \cite{kuehnast2004better} present a detailed study of favor exchange networks in the Kyrgyz Republic.\footnote{Their description of these relationships are in line with \cite{ledeneva1998russia}: ``\ldots informal social networks
were the most important mechanisms to get things done, obtaining access to ``deficit'' goods and services, acquiring accurate information about events and opportunities, circumventing
regulations and, in combination with bribes, gaining access to elite education, quality health care, and
positions of power. This network-based economy of reciprocal favors... was an important feature of the centralized socialist economy that helped people compensate for failures of the state.  ... the relatively egalitarian conditions of Soviet society enabled most people to establish
far-reaching networks.} Interestingly, the authors argue that the egalitarian conditions played a role in the formation of relationships and the size of the networks.\footnote{Indeed, in my model, when players are homogeneous, any two players can have a favor exchange relationship in a stable network.} However, the dissolution of the Soviet Union and the increased inequality that came with it had important effects on the structure of these social networks. Although social networks continue to be an integral part of everyday life in post-socialist Kyrgyz society,  \cite{kuehnast2004better} emphasize some important changes these networks have experienced:
\begin{enumerate}
    \item The size of networks and frequency of social encounters have significantly decreased
among the poor, leading to greater economic, geographic, and social isolation.
Simultaneously, the non-poor have become more reluctant to provide support to
poor relatives.
    
    \item Connections are still the main currency for gaining access to public services, jobs,
and higher education. However, the non-poor are able to use cash to supplement or
even substitute for connections.

    \item There is increasing differentiation in the form and function of social networks of the
poor and the non-poor. Polarization of these networks reflects increasing
socioeconomic stratification of the population.
\end{enumerate}

\subsection{Heterogeneous Players and Stratification of Networks}
Motivated by these observations, I consider a case where the rich and the poor have different costs to provide a favor, $c_r$ for the rich and $c_p$ for the poor. $N_r$ denotes the set of rich players. The rich have more resources at their disposal. They can use transfers ($\tilde N = N_r$) and can grant favors easier: $c_r<c_p$.  $B^*(c) \equiv B(\alpha,p,v,c,\delta)$ denotes the cooperation bound of a society with cost $c$. Given $c_p$, let $\tilde c_r = \inf \{c_r \in [0,c_p]: B^*(c_r) = B^*(c_p)\}$ denote the minimum cost level under which rich and poor players can support the same number of relationships.\footnote{$\tilde c_r$ is characterized in the proof of Proposition \ref{hetcostprop}.} The following proposition shows that whenever the inequality is high enough, the network is stratified.

\begin{prop}\label{hetcostprop}
The following statements are true:
\begin{itemize}
    \item   If $c_r > \tilde c_r$, then all networks where all players have $B^*$ links are strongly stable with transfers.
    \item If $c_r < \tilde c_r$ and $g$ is strongly stable with transfers, at $g$, all rich players have $B^*(c_r)$ links and none are linked with a poor player, who have at most $B^*(c_p)$ links. 
    \item If $c_r < \tilde c_r$, in any strongly stable network, as $N \to \infty$, the fraction of rich players who have $B^*(c_r)$ links, the fraction of poor players who have $B^*(c_p)$ links and the fraction of players who only have links with their own group goes to $1$.
\end{itemize}
\end{prop}

When $c_r \geq \tilde c_r$, the inequality is low that both types of players can sustain the same number of relationships, and heterogeneity does not have an effect on stable networks. However, when $c_r < \tilde c_r$, the inequality is high and lower relative costs allow rich players to sustain a larger network. This causes the poor players to be worse off through three different channels. First, the higher cost of favors directly reduces the benefits of favor exchange compared to rich players. Second, these higher costs decrease the number of relationships that poor players can support compared to rich players. With fewer relationships, favors are performed less frequently for the poor. Third, poor players are linked to other poor players who have fewer relationships and depend more on each of their neighbors. Since it is better to have players with more connections, the networks of poor players are not only smaller, but also composed of worse partners compared to rich players. These final two channels are direct consequences of substitutability and would not be obtained in a model with monopolistic cooperation. Moreover, they illustrate how inequality can result in the stratification of networks, which can further exacerbate the inequality in society when relationships are substitutable.

\begin{figure}
\caption{Strongly Stable Networks Before and After Heterogeneity}
\tikzstyle{mynode}=[thick,draw=blue,fill=blue!20,circle,minimum size=15]
\tikzstyle{mynode2}=[thick,draw=red,fill=red!20,circle,minimum size=15]
\tikzstyle{mynode3}=[thick,draw=teal,fill=teal!20,circle,minimum size=15]
\tikzstyle{oldnode}=[circle,draw,fill=black]
    \centering
\scalebox{1}{
\centering
\begin{tikzpicture}
\tiny
    \node[mynode] at (0, 1)   (r1) {};
    \node[mynode] at (0.8, 3.4)   (r2) {};
    \node[mynode] at (2, 1.7)   (r3) {};
    \node[mynode] at (-2.4, 2.7)   (r4) {};
    \node[mynode] at (-2, 4)   (r5) {};
    \node[mynode] at (0, 4)   (r6) {};
    \node[mynode] at (-4.2, 3)   (r7) {};
    \node[mynode] at (-3, 1)   (r8) {};
    \draw[] (r1) node[above,xshift=1cm] {} -- (r8);
    \draw[] (r4) node[above,xshift=1cm] {} -- (r7);
    \draw[] (r7) node[above,xshift=1cm] {} -- (r8);
    \draw[] (r6) node[above,xshift=1cm] {} -- (r2);
   \node[mynode2] at (-3, 1.8)   (p1) {};
    \node[mynode2] at (-5, 3)   (p2) {};
    \node[mynode2] at (-3, 4.5)   (p3) {};
    \node[mynode2] at (-4, 4)   (p4) {};
    \node[mynode2] at (-1.5, 0)   (p5) {};
    \node[mynode2] at (1.5, 0.5)   (p6) {};
    \node[mynode2] at (1.7, 2.8)   (p7) {};
    \node[mynode2] at (0, 2.5)   (p8) {};
    \draw[] (p1) node[above,xshift=1cm] {} -- (p2);
    \draw[] (p2) node[above,xshift=1cm] {} -- (p4);
    \draw[] (p5) node[above,xshift=1cm] {} -- (p6);
    \draw[] (p5) node[above,xshift=1cm] {} -- (p8);
    \draw[] (r5) node[above,xshift=1cm] {} -- (p3);
    \draw[] (r3) node[above,xshift=1cm] {} -- (p6);
    \draw[] (r5) node[above,xshift=1cm] {} -- (p4);
    \draw[] (r2) node[above,xshift=1cm] {} -- (p7);
    \draw[] (r4) node[above,xshift=1cm] {} -- (p3);
    \draw[] (r6) node[above,xshift=1cm] {} -- (p1);
    \draw[] (r3) node[above,xshift=1cm] {} -- (p8);
    \draw[] (r1) node[above,xshift=1cm] {} -- (p7);
\end{tikzpicture}
\begin{tikzpicture}
\tiny
    \node[mynode] at (0, 1)   (r1) {};
    \node[mynode] at (0.8, 3.4)   (r2) {};
    \node[mynode] at (2, 1.7)   (r3) {};
    \node[mynode] at (-2.4, 2.7)   (r4) {};
    \node[mynode] at (-2, 4)   (r5) {};
    \node[mynode] at (0, 4)   (r6) {};
    \node[mynode] at (-4.2, 3)   (r7) {};
    \node[mynode] at (-3, 1)   (r8) {};
    \draw[] (r1) node[above,xshift=1cm] {} -- (r2);
    \draw[] (r1) node[above,xshift=1cm] {} -- (r3);
    \draw[] (r1) node[above,xshift=1cm] {} -- (r7);
    \draw[] (r1) node[above,xshift=1cm] {} -- (r8);
    \draw[] (r2) node[above,xshift=1cm] {} -- (r3);
    \draw[] (r2) node[above,xshift=1cm] {} -- (r4);
    \draw[] (r2) node[above,xshift=1cm] {} -- (r8);
    \draw[] (r3) node[above,xshift=1cm] {} -- (r4);
    \draw[] (r3) node[above,xshift=1cm] {} -- (r5);
    \draw[] (r4) node[above,xshift=1cm] {} -- (r5);
    \draw[] (r4) node[above,xshift=1cm] {} -- (r6);
    \draw[] (r5) node[above,xshift=1cm] {} -- (r6);
    \draw[] (r5) node[above,xshift=1cm] {} -- (r7);
    \draw[] (r6) node[above,xshift=1cm] {} -- (r7);
    \draw[] (r6) node[above,xshift=1cm] {} -- (r8);
    \draw[] (r7) node[above,xshift=1cm] {} -- (r8);
   \node[mynode2] at (-3, 1.8)   (p1) {};
    \node[mynode2] at (-5, 3)   (p2) {};
    \node[mynode2] at (-3, 4.5)   (p3) {};
    \node[mynode2] at (-4, 4)   (p4) {};
    \node[mynode2] at (-1.5, 0)   (p5) {};
    \node[mynode2] at (1.5, 0.5)   (p6) {};
    \node[mynode2] at (1.7, 2.8)   (p7) {};
    \node[mynode2] at (0, 2.5)   (p8) {};
    \draw[] (p1) node[above,xshift=1cm] {} -- (p2);
    \draw[] (p2) node[above,xshift=1cm] {} -- (p4);
    \draw[] (p3) node[above,xshift=1cm] {} -- (p4);
    \draw[] (p5) node[above,xshift=1cm] {} -- (p6);
    \draw[] (p5) node[above,xshift=1cm] {} -- (p8);
    \draw[] (p6) node[above,xshift=1cm] {} -- (p8);
    \draw[] (p1) node[above,xshift=1cm] {} -- (p3);
\end{tikzpicture}
}
\floatfoot{\textit{Note:}  Parameter values: $\delta = 0.95$, $\alpha = 0.1$, $p=0.1$, $v=9$, $\gamma =0$. LHS is a strongly stable network when all players have costs $c = 1.3$, with $B^*(c) = 2$. RHS is a strongly stable network where red (poor) players have costs $c_p = 1.3$, but blue (rich) players have $c_r = 1$, with $B^*(c_r) = 4$. }
    \label{fig:stratification}
\end{figure}
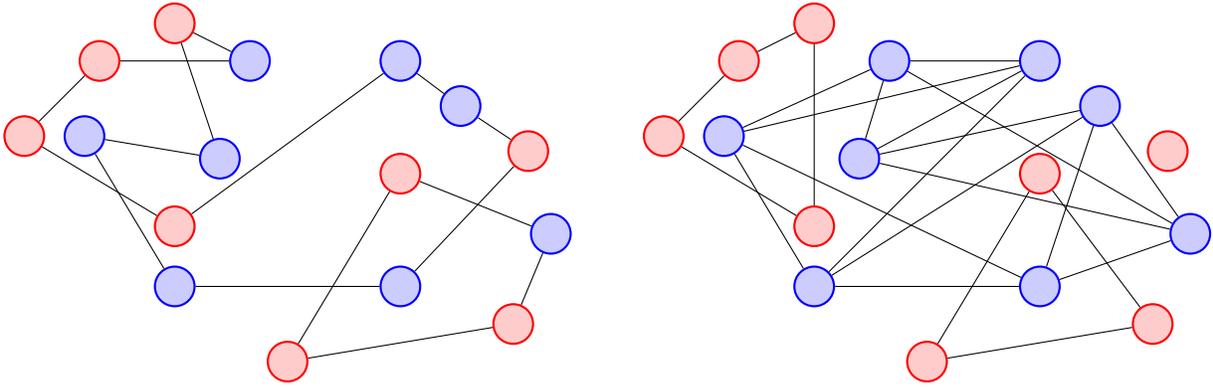

The predictions in Proposition \ref{hetcostprop} are in line with the main observations of \cite{kuehnast2004better}. During the Soviet era, the society was egalitarian and most individuals had networks where they exchanged favors, which is the equilibrium in the homogeneous case. However, following the collapse, society becomes more unequal, social networks become more polarized, and the size of the networks of the poor has decreased, exhibiting the changes predicted by Proposition \ref{hetcostprop}. The implications of the proposition are illustrated in Figure \ref{fig:stratification}. The left pane shows a strongly stable favor exchange network when players are homogeneous with costs $c=1.3$ and $B^*(c) = 2$. In this network, all players have $2$ relationships, and any such network would be strongly stable. In the right pane, red (poor) players still have cost $c_p = 1.3$ but blue (rich) players have $c_r = 1$, which is lower than $\tilde c_r \approx 1.19$. In this case, $B^*(c_r) = 4$ the plotted favor exchange network is strongly stable with transfers. In this network (and all other strongly stable networks), all rich players have exactly $4$ links, and they are only linked to each other. Poor players lose all links with rich players but may form new links with other poor players, even though not all are guaranteed to even reach $2$ links in strongly stable networks.\footnote{For example, the poor player who is in the top right of the graph has no links in this strongly stable network.}

\section{Alternative Enforcement Mechanisms}\label{communityenforcementsection}
In this section, I first describe how and why two contemporaneous trader groups (Maghribi and Genoese) have adopted two different enforcement mechanisms. Second, I study community enforcement and show that when effective, community enforcement crowds out any bilateral relationships, which explains the lack of relationships between foreign traders and Maghribi traders, who relied on community enforcement. Finally, I introduce costs for maintaining communication in a community and establishing a legal system, characterize the efficient mechanisms, and relate it to the choices made by Maghribi and Genoese traders.

\subsection{Maghribi and Genoese Traders}\label{maghribigenoese:background}

Maghribi traders are a group of Jewish traders who lived in the Mediterranean in the eleventh century and used a reputation-based institution to deal with the contractual problems inherent in merchant-agent transactions (see \cite{greif2006institutions} for a detailed description of the institutions Maghribi traders have used). Maghribi traders mainly relied on a multilateral reputation mechanism, where traders share information about each others' behavior with other traders and use a collective punishment strategy of not trading with agents who have previously cheated others. However, another group operating in the same era, the Genoese traders, have adopted a different enforcement mechanism that does not rely on collective punishments. \cite{greif1994cultural} summarizes these differences as follows:

\begin{quote}
Collectivist cultural beliefs were a focal point among the Maghribis, and individualist cultural beliefs were a focal point among the Genoese. Does the historical evidence indicate the existence of the related societal organizations? That is, was there high investment in information and collective punishment among the Maghribis and low investment in information and individualist punishment among the Genoese? The historical evidence indicates that the Maghribis invested in sharing information and the Genoese did not. Each Maghribi corresponded with many other Maghribi traders by sending informative letters to them with the latest available commercial information and ``gossip'', including whatever transpired in agency relations among other Maghribis. Important business dealings were conducted in public, and the names of the witnesses were widely publicized.

Although, most likely, not every Maghribi trader was familiar with all the others, belonging to the Maghribis was easily verifiable through common acquaintances, an extensive network of communication, a common religion, and a common language.
\end{quote}

Unlike Maghribi traders, Genoese traders have relied on a legal system and a bilateral reputation mechanism. This mechanism is described in \cite{greif2006institutions} as:
\begin{quote}
Thus for agency relations to be established
in an individualistic society, an external mechanism such as a legal system
backed by the state is needed to limit agents’ ability to embezzle
merchants' capital. A legal system complements an institution based on
individualistic cultural beliefs; it does not replace the associated bilateral
reputation institution. Where a legal system has only a limited ability to
restrict cheating (e.g., from misreporting profit expenses), a reputation
mechanism still has to be used. The extensive writing of agency contracts
suggests that this was indeed the case among the Genoese.
\end{quote}
Therefore, Genoese traders enforced cooperation through the possibility of court punishments and the threat of losing the bilateral relationship.\footnote{``\ldots among the Genoese an agent was induced to be honest by the fear that, if he was not, his relations with a particular merchant (who could have represented a family or even a clan) would be terminated \citep{greif2002economic}. ''} Moreover, \cite{greif1989reputation}  notes that Maghribi traders retained their separate identity within the Jewish communities they lived in while they were active in trade, but integrated within the community once they had to abandon trade due to political reasons. He attributes the preservation of this separate identity to its ability to provide ``a network for the transmission of information that facilitated agency relations.'' 

\cite{greif1989reputation} presents another important empirical regularity. He emphasizes that ``Evidence of business association between Maghribi traders and non- Maghribi traders (Jewish or Muslim) is rare'' and that their coalition is closed to outsiders. He attributes this to the fact that relatively higher short-run gains Maghribi traders obtain from cheating a foreign trader. In the next section, I show that when cooperation is substitutable and community enforcement enhances cooperation, it crowds out the formation of bilateral relationships, explaining the lack of relationships between Maghribi traders and outsiders.

\subsection{Community Enforcement}

Let $\Phi = \{\phi_1,\ldots,\phi_n\}$ denote a partition of the players. Each $\phi \in \Phi$ denotes a \textit{community}. I use $\phi(i)$ to denote the community of a player $i$. A community $\phi \in \Phi$ is a \textit{large community} if $|\phi| > B^*$ and is a \textit{small community} otherwise.  $\overline{\Phi} \subseteq \Phi$ denotes the set of large communities. A community is interpreted as a group of players who have invested in monitoring and/or communication so that information about the interactions between community members can be monitored and credibly communicated within the community. 

Formally, $\mathcal{\tilde H}$ denotes the set of all \textit{community histories} and let $\tilde h_{t,ij}$ denote the history between $i$ and $j$. If $\phi(i) = \phi(j)$, then $\tilde h_{t,ij}$ includes the periods and action profiles of all previous interactions between $i$, $j$ and any $k \in \phi(i)$. Otherwise, $\tilde h_{t,ij}$ only includes the periods and action profiles of all past interactions between $i$ and $j$, in other words, it is the bilateral history between $i$ and $j$. A society is described by $(N,\alpha,M,G,\delta,\Phi)$.\footnote{A community can be a singleton, which means that the player is not part of any community. A special case of this model where each $\phi$ is singleton corresponds to the baseline model.}

Definitions \ref{bilateralmeasuredef}, \ref{bilateraleqdefn} and \ref{defn:stable} are updated by replacing $h_{t,ij}$ with $\tilde h_{t,ij}$ and changing the word bilateral to community. I say that a network $g$ is \textit{community stable} if there exists $(g,\sigma_g)$ that is a stable community equilibrium. 

The following proposition shows that small communities cannot increase cooperation compared to bilateral enforcement and whenever community enforcement increases cooperation, it prevents the establishment of bilateral relationships.

\begin{prop}\label{prop:exogenouscom}
    Suppose that $g$ is a community stable network. Then the following statements are true:
    \begin{itemize}
        \item If $\phi(i)$ is a small community, then $d_i(g) \leq B^*$.
        \item If $d_i(g) > B^*$, then $\phi(i)$ is a large community, and $ij \in g$ only if $j \in \phi(i)$.
    \end{itemize}
\end{prop}

To gain intuition for this result, suppose that $i$ and $j$ are linked but not in the same community. Then their relationship must be enforced bilaterally. This implies that neither player can sustain more than $B^*$ relationships, since in that case they will strictly prefer not to provide the favor and lose the relationship. Next, suppose that $d_i(g) > B^*$. Then $i$ is already cooperating with more than $B^*$ players, and losing a particular relationship is much less important for $i$ compared to another player who has fewer than $B^*$ relationships. Therefore, $i$ will prefer not to provide a favor to any $j$ who is not in the same community, as this could only cause the loss of a relationship with $j$ and would not affect any other relationship. Thus, all neighbors of $i$ at $g$ must belong to the same community as $i$. This result offers an explanation for the lack of relationships between Maghribi traders and traders from other groups.

Moreover, this prediction is a direct consequence of substitutability of favors and would not be obtained under monopolistic cooperation.  If cooperation is monopolistic, the marginal benefit of establishing a bilateral relationship with a player outside the community does not depend on cooperation within the community, and this relationship may be sustainable, regardless of the size of the community. Proposition \ref{prop:exogenouscom} also shows that communities must be large enough to have an effect on the level of cooperation. Any player whose community is not large enough needs bilaterally enforced relationships and cannot have more than $B^*$ links.



\subsection{Efficiency of Different Enforcement Mechanisms}

In this section, I will formally define and compare three different mechanisms that can be used to sustain cooperation: pure bilateral enforcement, community enforcement, and legal enforcement.

\paragraph{Pure Bilateral Enforcement} Pure bilateral enforcement represents cases where cooperation cannot be enforced by a legal system or multilateral punishments. As outlined in Section \ref{blat:background}, many blat relationships and favor exchange relationships between individuals in modern societies fall under this category. This is the simplest enforcement mechanism in which players play the favor exchange game with $\gamma = 0$ and cooperation is sustained only by the threat of losing the relationship. The expected payoff per period under pure bilateral enforcement is given by plugging in $d_i(g) = B^*$ in Equation \ref{eq:expectedpayoff}.

\paragraph{Legal Enforcement}  In legal enforcement, players play the favor exchange game with $\gamma \geq 0$  and each player pays $C(\gamma)$ per period to maintain this legal system. $C$ is convex, $C(\gamma) > 0$ for all $\gamma$ and $\lim_{\gamma \to c} C(\gamma) = \infty$. Legal enforcement is a good model for Genoese traders and modern firms, especially when the quality of goods and services is important and not easily verifiable.

We can interpret legal enforcement in multiple ways. First, if the courts are already present, $C(\gamma)$ can represent the costs associated with making contractual agreements that could be enforced in courts.\footnote{Indeed, \cite{greif1994cultural} discusses that Genoese traders used documents called ``bill of landing'' to keep official records of their transactions, while Maghribi traders did not as they have solved this problem through their informal enforcement mechanisms.}  $C(\gamma)$ can also represent the per-player costs associated with maintaining the legal system.\footnote{In Appendix \ref{sec:popsize}, I modify the per-period cost to $C(\gamma)/N$, where $N$ is the population size and show that legal enforcement is optimal in large societies.} The efficient networks under legal enforcement is characterized in the following corollary, which follows from Proposition \ref{csprop} and Corollaries \ref{cor1}   and \ref{transfercor}.

\begin{cor}\label{legalcharacterization:cor}
For each $\gamma$, there is a $B^*(\gamma)$ such that a network is constrained efficient if and only if all players have $B^*(\gamma)$ links. $B^*(\gamma)$ is increasing in $\gamma$. Moreover, these are the only strongly stable networks with transfers.
\end{cor}

Let $g_{\gamma}$ denote the network in which all players have $B^*(\gamma)$ relationships, where all players obtain a per period payoff of $u(g_{\gamma}) \equiv u_i(g_{\gamma})$. Then the per period payoff of a player under legal enforcement, which is attained with the network $g_{\gamma}$, is $U(\gamma) \equiv u(g_{\gamma}) - C(\gamma)$. The optimal level of legal enforcement is given by $\gamma^* = \arg \max_{\gamma \in [0,c]} U(\gamma)$.\footnote{$\gamma^*$ is characterized in the proof of Proposition \ref{optimalcommunity:prop}.}

\paragraph{Community Enforcement} In community enforcement, each player belongs to a community and is linked to all members of their community. Players play the favor exchange game with $\gamma = 0$. Any player $i$ pays a per period cost $\kappa$ for each $j$ with $j \in \phi(i)$, which denotes the cost of maintaining links that allow for monitoring and communication channels.

Let $U(|\phi|,\kappa)$ denote the expected payoff per period of a member of a community of size $|\phi|$, including the costs associated with maintaining community links. 

\begin{prop}\label{optimalcommunity:prop}
There exists a $B(\kappa)$ such that $U(|\phi|,\kappa)$ is maximized at $|\phi| = B(\kappa)$.  $B(\kappa)$ is decreasing in $\kappa$. There are $\kappa_P$ and $\kappa_L$ such that:
\begin{itemize}
    \item Community enforcement with optimal community size $B(\kappa)$ results in a higher payoff than pure bilateral enforcement if and only if $\kappa \leq \kappa_P$. Moreover, whenever $\kappa \leq \kappa_P$, $B(\kappa) > B^*$.
    \item Community enforcement with optimal community size $B(\kappa)$ results in a higher payoff than optimal legal enforcement if and only if $\kappa \leq \kappa_L$.
\end{itemize}
\end{prop}

The optimal community size is a direct consequence of substitutability. When cooperation is enforced by communities, the (per-period) cost of adding a new member is $\kappa$, while the benefit decreases as the community grows. Therefore, it is optimal for players to organize themselves as communities of $B(\kappa)$. For a community to increase the payoff of its members, it must be larger than the number of relationships that can be sustained by pure bilateral enforcement, which happens when the cost of maintaining the community links is low enough. 

The second part of the proposition echoes Greif's observation about collectivism and individualism being a determinant of these different choices made by different groups of traders. Maghribi traders share a common religion and language, belong to a collectivistic culture, lived in a smaller society (which would correspond to lower $\kappa$), and did not integrate into the larger Jewish community in which they live (possibly to keep facilitate credible communication and $\kappa$ low) as long as they are active in trade. These features led to the adoption of community enforcement. On the contrary, Genoese traders lived in a larger, individualistic society, which precluded the adoption of community enforcement and paved the way for the development of a legal system. 

\section{Conclusion}
This paper introduces a novel framework to study favor exchange that allows favors to be substitutable. When favors are substitutable, the network structure becomes important in determining how frequently individuals interact with each other and affects the value of their relationship. By focusing on this new role of the network, I uncover an important relationship between substitutability of relationships and bilateral enforcement, and show that when players can rely on the rest of their network after losing a relationship, intermediate levels of cooperation are observed under bilateral enforcement. 

I extend the model to allow for transfers, heterogeneous players, and community enforcement. First, I apply the results to favor exchange networks in the Soviet Union. Motivated by studies on changes after the breakup of the Soviet Union, I study a society which is divided into two groups, rich and poor. The predictions of the model match the consequences emphasized in these studies: The networks of the rich grow and consist of other rich individuals while the networks of the poor contract and consist of other poor individuals. Second, I study the efficiency of different enforcement mechanisms in which players can invest in communication and use multilateral punishments or a legal system that punishes players who deviate. I show that community enforcement crowds out bilateral enforcement, which offers an explanation for the lack of relationships between Maghribi traders who practice community enforcement and outsiders. I also show that community enforcement is optimal when communication channels are easy to maintain, which is the case for a tightly knit society like the Maghribi traders, but not for the individualistic and larger society of Genoese traders, echoing the mechanisms adopted by Genoese and Maghribi traders described in \cite{greif1994cultural}.


\bibliographystyle{jfe}
\bibliography{bib-matching}
\vspace{-1em}
\begin{appendices}
\vspace{-1em}
\section{Preliminary Results}
\vspace{-0.5em}
I first prove an important Lemma about the beliefs in stable bilateral equilibria. I will write $\sigma_i(h_{t,i},j) = X$ if $\sigma_i$ puts probability $1$ to the action $X$. 
\begin{lem}\label{lem:allfavorsareprovided}
    Suppose that $g,\sigma$ is a stable bilateral equilibrium, $ij \in g$ and at $h_{t,i}$, both $i$ and $j$ provided all favors to each other and the link $ij$ is kept. Then $\sigma_i(h_{t,i},j) = (K,C)$.
\end{lem}
\begin{proof}
    For a contradiction, suppose that  $\sigma_i(h_{t,i},j) \neq (K,C)$ for some strictly positive probability. Let $p_R$ denote the probability $i$ plays $R$ and $p_D$ denote the probability $i$ plays $(K,D)$.
    
    Let $h_t$ denote any history that agrees with $h_{t,ij}$ and such that all favors are provided and all links are kept between all players (thus $h_{t,ij}$ is the bilateral history between $i$ and $j$ at $h_t$). Observe that under $\sigma$, $h_t$ has strictly positive probability, say $\beta$. If $p_R > 0$, then with probability $\beta p_R$, a link is not kept at $\sigma$, which means that $g,\sigma$ is not a stable bilateral equilibrium. If $p_R = 0$, then it must be that $p_D > 0$. Let $\beta_{ij}>0$ denote the probability that $j$ asks a favor from $i$ at $h_t$. Then $i$ refuses to provide the favor with probability $\beta \beta_{ij} p_D > 0$, which means that $g,\sigma$ is not a stable bilateral equilibrium. This proves the result.
\end{proof}
\begin{lem}\label{lem:offpathbeliefs}
    If $(g,\sigma,\mu)$ is a stable bilateral equilibrium, then at any history, any player believes that all favors are provided and all links are kept apart from the ones that he has already observed otherwise.
\end{lem}
\begin{proof}
    Take a player $i$ and history $h_{t,i}$. Let $H_t$ denote the set of all histories $h'_t$ such that  $h'_{t,i} = h_{t,i}$. Take any history $h'_{t} \in H_t$ such that there is a deviation between $j$ and $k$ with $i \not \in \{j,k\}$. Note that for each $t$, there are finitely many such $h'_t$. 

     Let $t' < t$ denote the period where first deviation occurred between $j$ and $k$ in $h'_t$. Moreover, As $(g,\sigma,\mu)$ is a stable bilateral equilibrium, the strategies of all players only depend to their bilateral history. Therefore that any $\sigma^k$ that converges to $\sigma$, the probability that there is a deviation between $j$ and $k$ at period $t'$ converges to $0$ as $k \to \infty$. Thus, $\lim_{k \to \infty} \mu^k(h'_t) = 0$. As there are finitely many such $h'_t$, the sum of probabilities attached to them also converges to zero. Therefore, at any history, any player believes that all favors are provided and all links are kept apart from the ones that he has already observed otherwise.
\end{proof}

\begin{cor}\label{cor:futurepayoff}
    Suppose that $(g,\sigma,\mu)$ is a stable bilateral equilibrium, $h_{t,i}$ is a private history where $i$ has not observed a deviation and $ij \in g$. Consider $\sigma'$ such that $\sigma'(h_{t,i},j) = \{K,D\}$ but is otherwise same as $\sigma$. If in period $t$, $j$ asks a favor from $i$ and $i$ is able to provide that favor, the difference between the continuation payoffs between $\sigma$ and $\sigma'$ is
    \begin{equation}
         c - \gamma + \frac{\delta}{1-\delta} (u_i(g) - u_i(g-ij)) 
    \end{equation}       
\end{cor}
\begin{proof}
    If  $j$ asks a favor from $i$ in period $t$, not providing the favor increases period $t$ utility by $c-\gamma$, but causes link $ij$ to break. By Lemma \ref{lem:offpathbeliefs}, $i$ believes that the rest of the network is unchanged under $\mu$, thus, by Lemma \ref{lem:allfavorsareprovided}, future payoff is given by $\frac{\delta}{1-\delta} (u_i(g) - u_i(g-ij))$.
\end{proof}
\section{Proofs}
\subsection{Proof of Proposition \ref{prop:networkeffect}}
To prove the first part, let $\hat U = [u_i(g+ij) - u_i(g)] - [u_i(g+ij+ik) - u_i(g+ij)]$ and $x= 1-p$. Note that
\begin{equation}
\begin{split}
        \hat U &= \alpha v [(x^{d_i(g)}-x^{d_i(g)+1}) - (x^{d_i(g)+1}-x^{d_i(g)+2})]\\
        &= \alpha v x^{d_i(g)} [1-x - x + x^2] =  \alpha v x^{d_i(g)} (1-x)^2 > 0
\end{split}
\end{equation}
To prove the second part, note that
\begin{equation}
\begin{split}
    u_i(g+jk) - u_i(g) =& \alpha c \left(\frac{1-(1-p)^{d_j(g)+1}}{d_j(g)+1} - \frac{1-(1-p)^{d_j(g)}}{d_j(g)}\right)\\ & + \mathbb I \left\{ ik \in d_i(g)\right\} \alpha c \left(\frac{1-(1-p)^{d_k(g)+1}}{d_k(g)+1} - \frac{1-(1-p)^{d_k(g)}}{d_k(g)}\right) > 0
    \end{split}
\end{equation}
where the inequality follows as $\frac{\partial}{\partial x} \frac{1-(1-p)^x}{x} = \frac{(1-p)^x(1-x \log(1-p))}{x^2} > 0$ whenever $(1-p) \in (0,1)$ and $x\geq1$.
\subsection{Proof of Proposition \ref{stabilityprop}}\label{stability proof}
\begin{lem}\label{properties}
Suppose that $ij \in g \cap \tilde g$, $d_i(\tilde g) \leq d_i(g)$ and $d_j(\tilde g) = d_j(g)$. Then $u_i(\tilde g) - u_i(\tilde g - ij) \geq u_i( g) - u_i( g - ij)$.
\end{lem}
\begin{proof}
\begin{equation}
        u_i(\tilde g) - u_i(\tilde g - ij) - (u_i( g) - u_i( g - ij)) = \alpha v \Big( (1-p)^{d_i(\tilde g)}-(1-p)^{d_i(g)} \Big) \geq 0
\end{equation}
where the equality follows from  $d_j(\tilde g) = d_j(g)$ and the inequality follows from $d_i(\tilde g) \leq d_i(g)$ and $p \in (0,1)$.
\end{proof}

To prove the if part of the proposition, let $g$ be a network such that all links are sustainable (\textit{i.e.,} the inequality in equation \ref{char} holds for all $ij \in g$). We will show that the following strategy-belief pairs constitute a bilateral equilibrium:
\begin{equation*}
    \sigma_i(h_{t,i},j) = \begin{cases}\{K,C\} &\text{ if } ij \in N_i(g_t) \text{ and all favors are performed between $i$ and $j$ at $h_{t,i}$} \\
    \{R,D\} &\text{ otherwise}
    \end{cases}
\end{equation*}

and $\mu_i^*$  puts positive probability only to the private histories of other players where all previous favors have been performed and links have been kept apart from the ones $i$ have already observed otherwise.

Let $\sigma_i$ denote another strategy that deviates from $\sigma^*_i$ at a history $h_{t,i}$. In what follows, we will show that, for each $j \neq i$, replacing  $\sigma_i(h_{t,i},j)$ with  $\sigma^*_i(h_{t,i},j)$ weakly increases the payoff of $i$, thus $\sigma_i$ is not a profitable deviation. Let $i \neq j$ denote another player.

\textbf{Case 1: There was a previous deviation in $h_{t,ij}$:} In this case, the link $\{ij\}$ cannot be active in period $t$ as $\sigma_j^*(h_{t,j},i) = \{R,D\}$. As a result, replacing $\sigma_i(h_{t,i},j)$ with $\{R,D\}$ does not change the utility of player $i$.

\textbf{Case 2:  There was not a previous deviation in $h_{t,ij}$:}

Let $\mathcal D(h_{t,i})$ denote the set of all $k$ such that there was a previous deviation in $h'_{t,ik}$, where $h'$ is a subhistory of $h$. Let $\tilde g$ denote the network obtained by removing all links $ik$ for $k \in \mathcal D(h_{t,i})$. Note that $\mu^*_i(h_{t,i})$ puts positive probability to histories where all favors are provided and all links kept between players $l$ and $l'$ unless $l \in \mathcal D(h_{t,i})$ and $l'=i$. Let $\omega_R$ denote the probability that $R$ is played under $\sigma_i(h_{t,i},j)$ and $\omega_D$ denote the probability that $\{K,D\}$ is played under $\sigma_i(h_{t,i},j)$. The difference in the expected continuation payoff at history $h_t$ between $\sigma^*$ and $\sigma$ is given by:
\begin{equation*}
    \begin{split}
           & \omega_R \left[ \frac{\delta }{1-\delta} \left(u_i(\tilde g)-  u_i(\tilde g-ij)\right)\right] +   \frac{\omega_D \alpha (1-(1-p)^{d_j(g)})}{d_j(\tilde g)} \left[-c + \gamma+ \frac{\delta }{1-\delta}\left( u_i(\tilde g)-  u_i(\tilde g-ij)\right)\right] \geq \\
         &  \omega_R \left[ \frac{\delta }{1-\delta} (u_i(g)- u_i(g-ij))\right] +     \frac{\omega_D \alpha (1-(1-p)^{d_j(g)})}{d_j(g)} \left[-c +\gamma + \frac{\delta }{1-\delta} (u_i(g)- u_i(g-ij))\right] \geq 0
    \end{split}
\end{equation*}

where the first inequality follows from  Lemma \ref{properties} and second from the fact that the link $ij$ is sustainable at $g$. Since this is true for all such $h_t$, at each history $\sigma$ does not agree with $\sigma^*$, it weakly lowers the expected payoff and $\sigma^*$ has higher expected payoff compared to $\sigma$.

To prove the only if part, suppose that the inequality in Equation \ref{char} doesn't hold for $ij \in g$ and $g$ is a stable network. Then there exists $\sigma$ such that $\sigma,g$ is a bilateral equilibrium where all favors are provided and all links are kept. As in any bilateral equilibrium all favors are performed on the equilibrium path, the expected utility under any bilateral equilibrium is same as the expected utility under $\sigma$. Let $\tilde \sigma$ denote an alternative strategy such that $\tilde \sigma_i(h_{0,i},j) = \{K,D\}$. By Corollary \ref{cor:futurepayoff}, the expected utility difference between $\tilde \sigma$ and $\sigma$ is
\begin{equation}
     \frac{ \alpha (1-(1-p)^{d_j(g)}) }{d_j(g)} \left[\frac{\delta}{1-\delta} u_i(g-ij) + c - \gamma - \frac{\delta }{1-\delta} u_i(g)\right]>0
\end{equation}

where the inequality follows as $ij$ is not sustainable. Thus $\tilde \sigma_i$ is a profitable deviation and $g$ is not a stable network. 

\vspace{-1em}
\subsection{Proof of Theorem \ref{T1}}\label{T1proof}

Let $i$ be a player who has $n$ neighbors (who all have $n$ neighbors) at $g$. For $g$ to be stable, from Proposition \ref{stabilityprop}, 
\begin{equation}
     -c +\gamma + \frac{\delta}{1-\delta}u_i(g) \geq \frac{\delta}{1-\delta}u_i(g - ij) \text{ for } j \in N_i(g)
\end{equation}
Rearranging, we obtain
\begin{equation}\label{stabilityconstraint}
\begin{split}
        - c + \gamma + \frac{\delta}{1-\delta}&\Big[\alpha v \Big( 1 - (1-p)^n \Big) - n \frac{\alpha}{n} c \Big( 1 - (1-p)^n \Big)\Big]\\
            & \geq 
         \frac{\delta}{1-\delta}\Big[\alpha v \Big(  1 - (1-p)^{n-1}  \Big) - (n-1) \frac{\alpha}{n} c \Big(  1 - (1-p)^n  \Big)\Big]       
\end{split}
\end{equation}

The equation highlights the trade-off between keeping the neighbor and incurring the cost and not incurring the cost but having less continuation payoff. Rearranging we obtain
\begin{equation}\label{stabilityconstraint2}
    -c + \gamma + \frac{\delta}{1-\delta}\left(\alpha v \left((1-p)^{n-1} - (1-p)^n\right) - \frac{\alpha c}{n} \left(1-(1-p)^n\right) \right) \geq 0
\end{equation}

Equation \ref{stabilityconstraint2} can be written this as a cutoff on $c$, denoting the highest possible value of $c$ to cooperate when a player has $n$ neighbors (who all have $n$ neighbors):
\begin{equation}\label{cutoffeq}
    c(n) = \frac{\delta \alpha v (1-p)^{n-1} p }{1-\delta + \frac{\alpha}{n} \delta [1 - (1-p)^n]} + \gamma
\end{equation}

First, notice that $\lim_{n \to \infty} c(n) = 0$, so the number of people a player can cooperate is bounded for $c > 0$. Now, we show $c(n) > c(n+1)$ for all $n$:
\begin{equation}
    \begin{split}
        c(n) &= \frac{\delta \alpha v (1-p)^{n-1} p}{1-\delta + \frac{\alpha}{n} \delta [1 - (1-p)^n]} + \gamma\\
        & = \frac{\delta \alpha v (1-p)^{n-1} p }{1-\delta + \frac{\alpha}{n} \delta [p + p (1-p) + p (1-p)^2 + \hdots + p (1-p)^{n-1}]} \frac{1-p}{1-p} + \gamma\\
        &> \frac{\delta \alpha v (1-p)^{n} p}{1-\delta + \frac{\alpha}{n} \delta [p (1-p) + p (1-p)^2 + \hdots + p (1-p)^{n}]} + \gamma\\
        &> \frac{\delta \alpha v (1-p)^{n} p}{1-\delta + \frac{\alpha}{n+1} \delta [p + p (1-p) + p (1-p)^2 + \hdots + p (1-p)^{n}]} + \gamma\\
        &> \frac{\delta \alpha v (1-p)^{n} p}{1-\delta + \frac{\alpha}{n+1} \delta [1 - (1-p)^{n+1}]} + \gamma = c(n+1)
    \end{split}
\end{equation}

Where first inequality holds as we only partially multiply the denominator with $(1-p)<1$ and second inequality is true as $p > (1-p)^k p$ for all $k \geq 1$. As $c(n)$ monotonically decreases to $0$, there is a $B^*$ such that $c(n) > c$ for all $n \leq B^*$ = $c(n) < c$ for all $n > B^*$.

To prove the first part, assume there is a  stable network $g$ where at least one player have more than $B^*$ links. Let $i$ be the player who has the most links at $g$. Then $d_i(g) = k > B^*$. Let $j \in N_i(g)$ and $N_j(g) = k' \leq k$. Then
\begin{equation*}
\begin{split}
       \frac{\delta}{1-\delta} (u_i(g) - u_i(g-ij))& - c + \gamma =  -c + \gamma + \frac{\delta}{1-\delta}\left(\alpha v \left((1-p)^{k-1} - (1-p)^k\right) - \frac{\alpha c}{k'} \left(1-(1-p)^{k'}\right) \right)\\
        &\leq -c + \gamma + \frac{\delta}{1-\delta}\left(\alpha v \left((1-p)^{k-1} - (1-p)^k\right) - \frac{\alpha c}{k} \left(1-(1-p)^{k}\right) \right)<0
\end{split}
\end{equation*}

where first inequality follows from $ \frac{\alpha c}{k'} \left(1-(1-p)^{k'}\right)$ is decreasing in $k'$ and second from the fact that $k > B^*$. Thus, by Proposition \ref{stabilityprop}, $g$ is not stable.

To prove the second part, assume at $g$ all players have exactly $B^*$ links. Then,  $-c + u_i(g) \geq u_i(g-ij) - \gamma$ for all $j$ and by Proposition \ref{stabilityprop} $g$ is a stable network.

\subsection{Proof of Theorem \ref{T2}}\label{T2proof}

Let $g$ denote the network where all players have $B^*$ links. Let $\{i,j\} = S$ and $g'$ denote a network obtainable from $g$ via deviations by $S$ that violates both conditions given in Definition \ref{strongstability}. 

\textbf{\underline{Case 1: Either $d_i(g') > B^*$ or $d_j(g') > B^*$.}} Observe that the only possible link that exists in $g'$ but does not exist in $g$ is $\{ij\}$. Thus, if $d_i(g') > B^*$, then  $d_i(g') = B^* + 1$ and  $d_j(g') \leq B^* + 1$. 

\begin{lem}\label{lem:notsustainable}
Suppose that  $d_i(g) = B^* + 1$ and  $d_j(g) \leq B^* + 1$. Then the link $ij$ is not sustainable at $g$.
\end{lem}
\begin{proof}
    For a contradiction, suppose that that $ij$ is sustainable. Then
    \begin{equation}
        u_i(g) - u_i(g-ij) = \alpha v (1-p)^{d_i(g)-1}p - \alpha c \left(\frac{1-(1-p)^{d_j(g)}}{d_j(g)}\right) \geq c - \gamma
    \end{equation}
    Moreover, 
    \begin{equation}\label{eq:notstable}
        \alpha v (1-p)^{B^*}p - \alpha c \left(\frac{1-(1-p)^{B^*+1}}{B^*+1}\right) \geq  \alpha v (1-p)^{d_i(g)-1}p - \alpha c \left(\frac{1-(1-p)^{d_j(g)}}{d_j(g)}\right) \geq c - \gamma
    \end{equation}
    where the first inequality holds has $d_j(g) \leq B^* + 1$ and the term $\frac{1-(1-p)^{d_j(g)}}{d_j(g)}$ is decreasing in $d_j(g)$. Let $g'$ denote the network where all players have $B^*+1$ neighbors. Equation \ref{eq:notstable} implies that all links in $g'$ are sustainable and by Proposition \ref{stabilityprop}, $g'$ is strongly stable, which is a contradiction to the definition of $B^*+1$.
\end{proof}

By Lemma \ref{lem:notsustainable}, the link $ij$ is not sustainable at $g'$, which is a contradiction.

\textbf{\underline{Case 2: $d_i(g') \leq B^*$ or $d_j(g') \leq B^*$.}} As the only possible link that exists in $g'$ but does not exist in $g$ is $\{ij\}$, for all $k \in N_i(g') \cup N_j(g')$, $d_k(g') \leq B^*$. 

\begin{lem}\label{lem:improvement}
    Suppose that $d_i(g) < B^*$, $d_j(g) = B^*-1$ and $ij \not \in g$. Then $u_i(g+ij) > u_i(g)$.
\end{lem}
\begin{proof}
Let $d_i(g)=n$. Then
    \begin{equation}
        \begin{split}
            u_i(g+ij) - u_i(g) = & \alpha v (1-p)^n p - \frac{\alpha c}{B^*} \left( 1-(1-p)^{B^*} \right) \\
            \geq & \alpha v (1-p)^{B^*} p - \frac{\alpha c}{B^*} \left( 1-(1-p)^{B^*} \right) \geq \frac{1-\delta}{\delta} (c-\gamma) > 0
        \end{split}
    \end{equation}
    where first  holds as $n < B^*$, second from the definition of $B^*$ and final by assumption that $c>\gamma$.
\end{proof}
\begin{lem}\label{lem:fullisbest}
    Suppose both $i$ and all $j \in N_i(g)$ has $B^*$ links at $g$. Moreover, both $i$ and all $j \in N_i(g')$ has (weakly) fewer than $B^*$ links at $g'$. Then $u_i(g) \geq u_i(g')$.
\end{lem}
\begin{proof}
     Consider $g''$ that is equivalent to $g'$ but all neighbors of $i$ has $B^*$ links under $g''$. As $u_i(g)$ is increasing in $d_k(g)$ for all $k \in N_i(g)$, $u_i(g'') \geq u_i(g')$. If $d_i(g'') = B^*$, then  $u_i(g'') = u_i(g)$, which shows that $u_i(g) \geq u_i(g')$. If $d_i(g'') < B^*$, then consider $g'''$ that adds $B^* - d_i(g'')$ neighbors to $i$ and all these neighbors have $B^*$ links (after adding their link with $i$). Note that $u_i(g) = u_i(g''')$. By Lemma \ref{lem:improvement}, $u_i(g''') > u_i(g'')$. This proves that $u_i(g) \geq u_i(g')$. 
\end{proof}

Applying Lemma \ref{lem:fullisbest} for $i$ and $j$, we have $u_i(g) \geq u_i(g')$ and $u_j(g) \geq u_j(g')$,  which is a contradiction. This proves the first part of the theorem. To prove the second part, I first prove the following lemma.

\begin{lem}\label{strongstabilitybound}
In any strongly stable network and $k < B^*$, there can be at most $k+1$ players with $k$ links.
\end{lem}
\begin{proof}
Let $g$ denote a strongly stable network. For a contradiction, assume for some $n < B^*$, there are $n'>n$ players with $n$ number of links. Then at least two of those players are not linked, let $i$ and $j$ denote those players. Then $g' = g \cup \{ij\}$  satisfies all conditions for violation of strong stability and thus $g$ is not strongly stable.
\end{proof}

As a result, at any strongly stable network, for any $n < B^*$ there can be at most $n$ players with $n$ links, which proves that total number of players with fewer than $B^*$ can be at most $\frac{(B^*)^2+B^*}{2}$, which is independent of $N$. To prove the corollary, note that, as $N \to \infty$, the fraction of players who do not attain the cooperation bound goes to $0$.

\subsection{Proof of Proposition \ref{csprop}}\label{csproof}

Comparative statics with respect to $c, \delta, v$ and $\gamma$ are immediate from differentiating the LHS of equation \ref{stabilityconstraint2} with respect to these variables. 


\subsection{Proofs of Proposition \ref{transferprop}}

     I first prove the following lemma. 
     \begin{lem}\label{lem:SSTreachingbound}
         If $g$ is strongly stable with transfers (SST) and $i \in \tilde N$, then $d_i(g) = B^*$. 
     \end{lem}
     \begin{proof}
              For a contradiction, suppose that $g$ is SST and there exists $i \in \tilde N$ with $d_i(g) < B^*$. Moreover, let $i$ be one of the lowest degree players in $\tilde N$ at $g$, in other words, for all $j \in \tilde N$, $d_i(g) \leq d_j(g)$. As $g$  is SST, it is (strongly) stable and $d_l(g) \leq B^*$ for all $l$.
     
     Take any $j  \not \in N_i(g)$ and $j \in \tilde N$.\footnote{Such a $j$ exists as $|\tilde N| \geq B^*$ and $d_i(g) < B^*$}. If $d_i(g) = d_j(g)$, then $i$ and $j$ violates strong stability with transfers at $g$ via $g+ij$ and $t=(t_i,t_j) = (0,0)$, which is a contradiction. As we chose $i$ as one of the lowest degree member of $\tilde N$, $d_i(g) \neq d_j(g)$ implies $d_j(g) > d_i(g)$.
     
     As $d_j(g) > d_i(g)$, there exists some $k \in N_j(g)$ and $k \not \in N_i(g)$. Consider the network $g' = g + ij - jk$, which is obtainable from $g$ via deviations by $\{i,j\}$. I will show that there exists $t = (t_i,t_j)$ such that $i,j$ violates SST at $g$ via $g'$.

     \textbf{\underline{Case 1: $d_k(g) \leq d_i(g')$.}} 
      Set $t_j = t_j = 0$. As transfers are zero, $\tilde u_i(g',j,t_i,t_j) = u_i(g')$ and $\tilde u_j(g',i,t_j,t_i) = u_j(g')$. As $d_k(g) \leq d_i(g')$, by Lemma \ref{lem:fullisbest}, $u_j(g') \geq u_j(g)$. Moreover,
      \begin{equation*}
      \begin{split}
            u_i(g') - u_i(g) &= \alpha v (1-p)^{d_i(g)} - \alpha c \frac{1-(1-p)^{d_j(g)}}{d_j(g)} \geq  \alpha v (1-p)^{d_i(g)} - \alpha c \frac{1-(1-p)^{d_i(g)}}{d_i(g)}\\
            &\geq  \alpha v (1-p)^{B^*} - \alpha c \frac{1-(1-p)^{B^*}}{B^*} \geq \frac{\delta}{1-\delta} > 0
      \end{split}
      \end{equation*}
      where the first inequality holds as $d_i(g) \leq d_j(g)$, second inequality holds as $d_i(g) \leq B^*$, third by definition of $B^*$ and fourth by our assumption that $c > \gamma$. Thus condition (ii) of a violation of SST are satisfied. To show that $ij$ is sustainable, first note that as $t=(0,0)$ and $d_i(g) \leq d_j(g) \leq B^*$, the sustainability condition for $i$ holds. Finally, as $g'-ij = g-jk$:
      \begin{equation}
          u_j(g') - u_j(g'-ij) \geq  u_j(g) - u_j(g-jk) \geq \frac{1-\delta}{\delta} (c-\gamma) 
      \end{equation}

      where the first inequality holds as $u_j(g') \geq u_j(g)$ and second holds as the link $kj$ is sustainable at $g$ (as $g$ was assumed to be SST). This shows that $\{i,j\}$ violates SST at $g$ via $g'$.

      \textbf{\underline{Case 2: $d_k(g) > d_i(g')$.}} If $d_k(g) > d_j(g)$, rename $j$ as $k$ and $k$ as $j$, otherwise do nothing. Then, we have $d_j(g) \geq d_k(g) > d_i(g')$, $jk \in g$, $ij \not \in g$ and $ik \not \in g$.  Set $t_j = 0$ and $t_i$ such that $\tilde u_j(g',i,0,t_i) = u_j(g)$, which implies condition (ii) in Definition \ref{strongstabilitytransfer} is satisfied for $j$ with equality. As $d_k(g) > d_i(g')$, $t_i > 0$.
      
      Moreover, as $u_j(g' - ij) = u_j(g-jk)$, we have $\tilde u_j(g',i,0,t_i) - u_j(g'-ij) =  u_j(g) -  u_j(g-jk)$, which implies Equation \ref{eq:sustainabilityj} is satisfied, as the link $jk$ was sustainable at $g$ (by stability of $g$).

      Writing down the utility difference for both players (see Footnote \ref{footnote:transfer} for the definition of $w$.):
      \begin{equation}
      \begin{split}
                    \tilde u_j(g',i,0,t_i) - u_j(g) &= - w(i,g') (c-t_i) + w(k,g) c\\
           \tilde u_i(g',j,t_i,0) - u_i(g) &= - w(i,g') t_i + \alpha (1-p)^{d_i(g)} p v - w(j,g) c
        \end{split}
      \end{equation}
      Summing both sides and plugging in $  \tilde u_j(g',i,0,t_i) - u_j(g) = 0$:
      \begin{equation}\label{eq:prop4-1}
    \begin{split}
                 \tilde u_i(g',j,t_i,0) - u_i(g)  &= - w(i,g') c + \alpha (1-p)^{d_i(g)} p v  + w(k,g) c - w(j,g) c\\
                 & \geq  - w(i,g') c + \alpha (1-p)^{d_i(g)} p v
    \end{split}
      \end{equation}
      where the inequality holds as $d_j(g) \geq d_k(g)$ implies $w(k,g) \geq w(j,g)$. Moreover, as $ u_j(g' - ij) =  u_j(g)$, we have $ \tilde u_i(g',j,t_i,0) - u_j(g) =  \tilde u_i(g',j,t_i,0) - u_j(g' - ij)$. Thus, showing RHS of Equation \ref{eq:prop4-1} is greater than $\frac{1-\delta}{\delta} (c-\delta) > 0$ implies the sustainability condition for $i$ and shows that $i$ is strictly better off at $g'$, which means that all conditions for the violation of SST will be satisfied. To see that, note 
      \begin{equation}
        \begin{split}
          - w(i,g') c + \alpha (1-p)^{d_i(g)} p v &= \alpha (1-p)^{d_i(g)} p v - \alpha c \frac{1-(1-p)^{d_i(g)+1}}{d_i(g)+1}\\
          & \geq  \alpha (1-p)^{B^*-1} p v -  \alpha c \frac{1-(1-p)^{B^*}}{B^*} \geq \frac{1-\delta}{\delta} (c-\gamma) > 0
        \end{split}
      \end{equation}
        where first inequality holds as $d_i(g) < B^*$ and second from the definition of $B^*$. Thus, $ij$ is sustainable at $g'$ with $t =(t_i,t_j)$, $\tilde u_i(g,j,t_i,t_j) - u_i(g-ij) > 0$ and   $\tilde u_j(g,i,t_j,t_i) - u_j(g-ij) \geq 0$, showing $g$ is not SST.
             \end{proof}

This finishes the proof of proposition. In Appendix \ref{sec:SSTnetworkexample} I also prove that the set of SST networks is not empty.
\vspace{-1em}
\subsection{Proof of Proposition \ref{monopolistic coop}}\label{monopolistic coop proof}
To prove the first part, assume  $c -\gamma \leq \dfrac{\delta \hat p (v-c)}{1-\delta}$. We will show the following strategy-belief pairs are a bilateral equilibrium under the complete network:
\begin{equation*}
    \sigma^*_i(h_{t,i},j) = \begin{cases}\{K,C\} &\text{ if } ij \in N_i(g_t) \text{ and all favors are performed between $i$ and $j$ at $h_{t,i}$} \\
    \{R,D\} &\text{ otherwise}
    \end{cases}
\end{equation*}

and $\mu_i^*$ puts positive probability only to the private histories of other players where all previous favors have been performed and links have been kept apart from the ones $i$ have already observed otherwise.

Let $\sigma_i$ denote another strategy that is different from $\sigma^*_i$. In what follows, we will show that, for each  $h_{t,i}$ such that $\sigma_i$ deviates from $\sigma^*_i$ and $j \neq i$, replacing  $\sigma_i(h_{t,i},j)$ with  $\sigma^*_i(h_{t,i},j)$ weakly increases the payoff of $i$, thus $\sigma_i$ is not a profitable deviation. Let $i \neq j$ denote another player.

\textbf{Case 1: There was a previous deviation in $h_{t,ij}$:} In this case, the link $\{ij\}$ cannot be active in period $t$ as $\sigma_j^*(h_{t,j},i) = \{R,D\}$. As a result, replacing $\sigma_i(h_{t,i},j)$ with $\{R,D\}$ does not change the utility of player $i$.

\textbf{Case 2:  There was not a previous deviation in $h_{t,ij}$:}

Let $\omega_R$ denote the probability that $R$ is played under $\sigma_i(h_{t,i},j)$ and let $\omega_D$ denote the probability that $\{K,D\}$ is played under $\sigma_i(h_{t,i},j)$. Note that the difference in the expected continuation payoff between $\sigma^*$ and $\sigma$ is given by
\begin{equation}
        \omega_R\left(\frac{1}{1-\delta}\frac{\alpha p}{N} (v-c)\right) +   \omega_D \frac{\alpha p}{N} \left( -c + \gamma + \frac{\delta}{1-\delta} \frac{\alpha p}{N} (v-c)  \right) > 0
\end{equation}

where the first term is positive since $v-c>0$ and second is positive from assumption above and the inequality follows. Thus, in both cases, $\sigma^*$ does better than $\sigma$ and $\sigma$ is not a profitable deviation, proving the first part of the result.

To prove the second part, assume  $c - \gamma> \dfrac{\delta \hat p (v-c)}{1-\delta}$ and that there is a stable network $g$ such that $ij \in g$. Thus, there is a bilateral equilibrium strategies $\sigma$ such that when all players follow $\sigma$, all favors are provided on the equilibrium path. Clearly, $\sigma_i$ assigns $\{K,C\}$ to both players in all $h_{t,ij}$ with no previous deviation, as otherwise $g$ would not be stable. Define $\sigma'$ by replacing $ \sigma_i(h_{0,i},j) = \{K,C\}$ with $\{K,D\}$ and leaving the rest of the strategy unchanged. The probability that $i$ requires favor $j$ is $\alpha/N$, and $j$ is able to provide that favor with probability $p$. Thus, by Corollary \ref{cor:futurepayoff} difference in expected continuation payoff between $\sigma$ and $\sigma'$ is given by:
\begin{equation}
   \frac{\alpha p}{N} \left( -c + \gamma + \frac{\delta}{1-\delta} \frac{\alpha p}{N} (v-c)  \right) = \hat p \left( -c + \gamma + \frac{\delta}{1-\delta} \hat p (v-c)   \right) < 0
   \end{equation}
  where the last inequality follows from our assumption above. As a result, $\sigma'$ is a profitable deviation and there cannot exits such an equilibrium. 
 
\subsection{Proof of Proposition \ref{hetcostprop}}
First, note that RHS of equation \ref{cutoffeq} is same for both rich and poor players. Thus, there exists $c$ such that equation \ref{cutoffeq} holds with equality when $n = B^*(c_p)+1$, which is $\tilde c_r$. To prove the first part, suppose that the inequality is small. The result follows from the same steps in Lemma \ref{lem:SSTnetworkexample} in Appendix \ref{sec:SSTnetworkexample}, replacing $c$ with $c_r$ if $i \in N_r$ and with $c_p$ if $i \in N_p$. 

To prove the second part, suppose that  the inequality is large and $\tilde N = N_r$.
\begin{lem}\label{lem:richlemma}
    If $i \in N_r$, then $d_i(g) = B^*(c_r)$.
\end{lem}
\begin{proof}
    Follows from replicating the steps in Lemma \ref{lem:SSTreachingbound}.
\end{proof}

\begin{lem}
    If $i \in N_p$, then $d_i(g) < B^*(c_r)$.
\end{lem}
\begin{proof}
    Suppose that there exists $i \in N_p$ with $d_i(g) \geq B^*(c_r)$. Without loss of generality, let $i$ be the poor player with highest number of links. Take any $k \in N_i(g)$. We have the following 
    \begin{equation}\label{eq:poorlinkunsustainable}
\begin{split}
            \alpha v (1-p)^{d_i(g)-1} p - & \alpha c \frac{1-(1-p)^{d_k(g)}}{d_k(g)}  \leq \alpha v (1-p)^{B^*(c_r)-1} p - \alpha c \frac{1-(1-p)^{d_k(g)}}{d_k(g)} \\ 
            &\leq    \alpha v (1-p)^{B^*(c_r)-1} p - \alpha c \frac{1-(1-p)^{B^*(c_r)}}{B^*(c_r)} < \frac{1-\delta}{\delta} (c_p -\gamma)
\end{split}
    \end{equation}
    where the first inequality holds as $d_i(g) \geq B^*(c_r)$, the second inequality holds as $d_k(g) \leq B^*(c_r)$ and third holds from the definition of $B^*(c_r)$ and the fact that $c_p > c_r$. Equation \ref{eq:poorlinkunsustainable} implies the link $ik$ is not sustainable, thus $g$ is not stable, which is a contradiction.
\end{proof}
    \begin{lem}\label{lem:fullstratification}
        If $i \in N_r$ and $j \in N_p$, then $ij \not \in g$.
    \end{lem}
    \begin{proof}
        Suppose that $ij \in g$. Take a $k \in N_r$ with $k \not \in N_i(g)$. Moreover, as $d_i(g) = d_k(g) = B^*(c_r)$, there exists $m$ such that $km \in N_k(g)$ with $m \not \in d_i(g)$. Consider $t=(0,0)$ and $g' = g - ij - km + ik$. We will show that this constitutes a violation of SST. First, as  $t=(0,0)$, $\tilde u_i(g',j,0,0) = u_i(g')$ and  $\tilde u_j(g',i,0,0) = u_j(g')$. Thus
        \begin{equation}
            u_i(g') - u_i(g) = \alpha c \left( \frac{1-(1-p)^{B^*(c_r)}}{B^*(c_r)} - \frac{1-(1-p)^{d_j(g)}}{d_j(g)} \right) > 0
        \end{equation}
where the inequality holds as $B^*(c_r) > d_j(g)$. Similarly, 
        \begin{equation}
            u_k(g') - u_k(g) = \alpha c \left( \frac{1-(1-p)^{B^*(c_r)}}{B^*(c_r)} - \frac{1-(1-p)^{d_m(g)}}{d_m(g)} \right) \geq 0
        \end{equation}
        where the inequality holds as $d_i(g') = B^*(c_r) \geq d_m(g)$. 
        Finally, observe that $u_k(g'-ik) = u_k(g-km)$ and $u_i(g'-ik) = u_i(g-ij)$. Thus, the sustainability of $ij$ at $g'$ is implied by the sustainability of $ij$ at $g$ (which holds as $g$ is assumed to be stable). This shows that $g$ is not SST, which is a contradiction.
    \end{proof}
    The following lemma proves the second part of the proposition.
    \begin{lem}
        If $i \in N_r$, then $d_i(g) = B^*(c_r)$. If  $i \in N_p$, then $d_i(g) \leq B^*(c_p)$
    \end{lem}
    \begin{proof}
        By Lemma \ref{lem:fullstratification}, at $g$, there is no link between players in $N_r$ and $N_p$. Let $g_p$ and $g_r$ denote the collection of links between players in $N_p$ and $N_r$. Then $g$ is stable if and only if $g_p$ is stable in $N_p$ and $g_r$ is stable in $N_r$. Thus, by Theorem \ref{T1}, if $i \in N_r$, $d_i(g) \leq B^*(c_r)$ and $i \in N_p$, $d_i(g) \leq B^*(c_p)$. From Lemma \ref{lem:richlemma}, we have   $d_i(g) = B^*(c_r)$, which proves the result.
    \end{proof}
    Third part of the proposition is an immediate from Proposition \ref{heterogeneity} in Appendix \ref{morehet}.

\subsection{Proof of Proposition \ref{prop:exogenouscom}} 
 \begin{lem}\label{lem:allfavorsareprovidedcommunity}
    Suppose that $g,\sigma$ is a community equilibrium, $ij \in g$ and at $\tilde h_{t,i}$, both $i$ and $j$ provided all favors to each other and the link $ij$ is kept at $\tilde h_{t,ij}$. Then $\sigma_i(\tilde h_{t,i},j) = (K,C)$.
\end{lem}
 \begin{proof}
     Follows from precisely same steps as in the proof of Lemma \ref{lem:allfavorsareprovided}.
 \end{proof}

 \begin{lem}\label{lem:offpathbeliefscommunity}
     Suppose that $g,\sigma$ is a community equilibrium, $ij \in g$ and $\phi(i) \neq \phi(j)$. Take a history $\tilde h_{t,i}$ such that $i$ has observed a deviation only in his bilateral history with $j$. Then $i$ believes that all favors are provided and all links are kept apart from what he has observed in his bilateral history with $j$.
 \end{lem}
 \begin{proof}
     Follows from precisely same steps as in the proof of Lemma \ref{lem:offpathbeliefs}.
 \end{proof}
 
Next, I prove the following lemma.

\begin{lem}\label{lem:onlycomlinks}
    Suppose that $g$ is a community stable network and $ij \in g$. If $\phi(i) \neq \phi(j)$, then $\max (d_i(g),d_j(g)) \leq B^*$.
\end{lem}
\begin{proof}

 For a contradiction, suppose that  $\max (d_i(g),d_j(g)) > B^*$. Without loss of generality, let $d_i(g) \geq d_j(g)$. As in any community equilibrium all favors are performed on the equilibrium path, the expected utility under any bilateral equilibrium is same for all $\sigma$. Let $\tilde \sigma$ denote an alternative strategy such that $\tilde \sigma_i(\tilde h_{0,i},j) = \{K,D\}$ and removes the link with $j$ whenever a favor is not performed in the past, but otherwise same as $\sigma$.

 From Lemmas \ref{lem:allfavorsareprovidedcommunity} and \ref{lem:offpathbeliefscommunity}, the expected utility difference between two strategies is given by.
\begin{equation}
      \frac{ \alpha (1-(1-p)^{d_j(g)})}{d_j(g)} \left[ \frac{\delta }{1-\delta} u_i(g) - c + \frac{\delta}{1-\delta} u_i(g-ij)\right] < 0
\end{equation}

where the inequality follows as $d_i(g) > B^*$ and  $d_i(g) \geq d_j(g)$. Thus $\tilde \sigma_i$ is a profitable deviation and $g$ is not a community stable network.
\end{proof}
To prove the first part, for a contradiction suppose $d_i(g) > B^*$ and $\phi(i)$ is a small community. As $\phi(i)$ is a small community,  there exists a $j$ such that $j \in d_i(g)$ and $j \not \in \phi(i)$. By Lemma \ref{lem:onlycomlinks}, this is a contradiction and the result follows.

To prove the second part, suppose that $d_i(g) > B^*$ and $i$ has a link with $j$ with $j \not \in \phi(i)$. This contradicts Lemma \ref{lem:onlycomlinks} and thus $g$ is not a community stable network.

\subsection{Proof of Proposition \ref{optimalcommunity:prop}}\label{optimalcommunity:proof}

First, note that
\begin{equation}
   U(n+1,\kappa) =\alpha v \Big( 1 - (1-p)^n \Big) - \alpha c \Big( 1 - (1-p)^n \Big) - n\kappa \equiv \beta(n) - n\kappa 
\end{equation}

Observe that $\dfrac{\partial^2}{\partial n^2}\beta(n) = -(1-p)^n \log^2(1-p) < 0$ for all $p \in (0,1)$. Thus $ U(n,\kappa)$ is strictly concave in $n$. Next, $\lim_{n \to \infty} \beta(n) = \alpha (v-c)$ and $\lim_{n \to \infty} n \kappa = - \infty$. Hence, $\lim_{n\to \infty}U(n,\kappa) = - \infty$. As $U(0,\kappa) = 0$, $U(n,\kappa)$ is either decreasing in $n \in [0,\infty)$ or $U(n,\kappa)$ has a unique interior optimum at some $n \in \mathbb R$. In the first case, the optimum is $0$, while in the second case $U$ is maximized at either the floor or ceiling of the interior optimum, which proves the first part. The fact that $B(\kappa)$ is decreasing in $\kappa$ is immediate.

To compare community enforcement with pure bilateral enforcement, note that when $\kappa$ is small, cooperation with any number of players can be supported at a very low cost, therefore community enforcement is optimal. Second, when $\kappa$ is large, clearly the cost of maintaining the community links dominates the benefit of cooperation, therefore bilateral enforcement is optimal. The following lemma shows that if community enforcement is optimal for a $\kappa$, then community enforcement is optimal under any $\kappa' < \kappa$.

\begin{lem}\label{lem:decreasinginkappa}
    $U(B(\kappa),\kappa)$ is decreasing in $\kappa$.
\end{lem}
\begin{proof}
    Let $B(\kappa)$ denote the optimal community size under $\kappa$. Note that 
\begin{equation}
    U(B(\kappa'),\kappa') \geq U(B(\kappa),\kappa') > U(B(\kappa),\kappa)
\end{equation}

where the first inequality holds from the optimality of $B(\kappa')$ under $\kappa'$ and second from the fact that $U$ is decreasing in $\kappa$. Thus community cooperation under $\kappa'$ is better than bilateral cooperation. 
\end{proof}

Therefore, there exists $\kappa_P$ such that community enforcement results in a higher payoff than pure bilateral enforcement if and only if $\kappa \leq \kappa_P$. Finally, if community cooperation is optimal under $\kappa$, then $B(\kappa) > B^*$ since cooperation of $B^*$ players can be sustained without community enforcement.

To prove the second part, note that the utility under legal enforcement is given by
\begin{equation}
  U(\gamma) = \alpha (v-c) \left(1-(1-p)^{B^*(\gamma)}\right) - C(\gamma)
\end{equation}

First, note that the first term is bounded above by $ \alpha (v-c) $. Second, as $\lim_{\gamma \to c} C(\gamma) = \infty$, there exists $\overline \gamma$ such that $U(\gamma) < 0$ for all $\gamma > \overline \gamma$. Let $\overline N = B^*(\overline \gamma)$. For each integer $n < \overline N$, let $\gamma_n$ denote the $\gamma$ value that satisfies the inequality in Equation \ref{stabilityconstraint2} with equality, which means that $\gamma_n$ is the lowest $\gamma$ such that players can sustain $n$ links. Take an $n^* \in \arg \max_{n < \overline N} U(\gamma_n)$, which is non-empty. Let $\gamma^* =\gamma_{n^*}$, which is the optimal level of legal punishment and $U(\gamma^*)$ is the highest per period utility players can obtain under legal enforcement. If $\kappa$ is small enough, then $n^* \kappa < C(\gamma^*)$, and community enforcement is more efficient than legal enforcement. Conversely, if $\kappa$ is large enough (\textit{e.g.}, $\kappa > \alpha v$), community enforcement is not optimal. Then Lemma \ref{lem:decreasinginkappa} implies that  there exists $\kappa_L$ such that community enforcement results in a higher payoff than pure bilateral enforcement if and only if $\kappa \leq \kappa_L$.

  
\section{Additional Results and Discussion}\label{app:hetagents}
\subsection{Bounded Cooperation under General Favor Provision Matrices}\label{generalm}

Let $F$ denote the finite set of different favor types and $M$ denote the favor provision probability matrix where $p_{if} \in 0 \cup [\underline p,1]$ for some $\underline p > 0$, in other words, the probability each player can provide each favor is either $0$, or bounded below by some number.

The following proposition shows that the bounded nature of cooperation is due to the substitutability of the favors and not symmetry of the fully substitutable model.

\begin{prop}\label{generalmprop}
For any $(N,\alpha,M,G_f,\delta,\underline p)$, there is a $B(\alpha,\underline p,G_f,\delta)$ such that if there exists $j$ with $d_j(g) > B(\alpha,\underline p,G_f,\delta)$ then $g$ is not stable.
\end{prop}
\begin{proof}
    Let $\hat u_i$ denote the expected per period utility of player $i$. Suppose that $\sigma,g$ is a stable bilateral equilibrium under beliefs $\mu$. For $ij \in g$, let $\omega(i,j,g)$ denote the probability  $j$ requests a favor from $i$ and $i$ is able to provide it at any given period under $g$.
    \begin{cor}
        Suppose that $ij \in g$. Consider $\sigma'$ such that $\sigma'(h_{0,i},j) = \{K,D\}$, and $\sigma'$ is same as $\sigma$ otherwise. The difference in continuation payoffs between $\sigma'$ and $\sigma$
        \begin{equation}
            \omega(i,j,g) \left[- c + \gamma + \frac{\delta}{1-\delta} (\hat u_i(g)) - \hat u_i(g-ij) \right]
        \end{equation}
    \end{cor}
    \begin{proof}
    First, unless $j$ requests a favor from $i$ and $i$ is able to provide it, both strategies are equivalent and gives $i$ the same expected payoff. 
    Suppose that at period $0$, $j$ requests a favor from $i$ and $i$ is able to provide it. Under $\sigma'$, $i$ plays $D$. By Lemma \ref{lem:offpathbeliefs}, $i$ believes that the rest of the network is unchanged under $\mu$, and thus by Lemma \ref{lem:allfavorsareprovided}, $i$ gets a payoff of $-\gamma + \frac{\delta}{1-\delta} \hat u_i(g-ij)$. Under $\sigma$, $j$ plays $C$, and by Lemma \ref{lem:offpathbeliefs} $i$ gets a payoff of $-c + \frac{\delta}{1-\delta} \hat u_i(g-ij)$.
    \end{proof}
    As $\sigma,g$ is a stable bilateral equilibrium, it must be that 
    \begin{equation}
         \frac{\delta}{1-\delta} (\hat u_i(g)) - \hat u_i(g-ij) \geq c - \gamma
    \end{equation}
    Suppose that at $g$, $i$ has $n_f$ links who can provide favor $f$ with positive probability. Let $\underline n = \min_{f} n_f$.
    \begin{equation}\label{eq:hetbenefit}
        \begin{split}
             \frac{\delta}{1-\delta} (\hat u_i(g)) - \hat u_i(g-ij) &\leq \frac{\delta}{1-\delta} \left( \sum_{f \in F} \alpha v \frac{1}{|F|} (1-\underline p)^{\underline n} p_{jf} \right)\\
             &\leq \frac{\delta}{1-\delta} \left( \sum_{f \in F} \alpha v \frac{1}{|F|} (1-\underline p)^{\underline n} \right)\\
        \end{split}
    \end{equation}
    where first inequality holds the probability that $j$ is pivotal is maximized when $n_f = \underline n$ and all neighbors of $i$ can provide each favor with probability $\underline p$. Second inequality holds as setting $p_{jf} = 1$ maximizes the term in the brackets. Note that RHS of Equation \ref{eq:hetbenefit} goes to $0$ as $\underline n$ goes to $\infty$. Then there exists $b^*$ such that 
    \begin{equation}
        \left( \sum_{f \in F} \alpha v \frac{1}{|F|} (1-\underline p)^{b^*} \right) \geq c- \gamma >   \left( \sum_{f \in F} \alpha v \frac{1}{|F|} (1-\underline p)^{b^*+1} \right)
    \end{equation}
    This implies the following lemma.
    \begin{lem}\label{lem:hetupperbound}
        If $g$ is stable and $ij \in g$, there is at least one favor type $f$ such that $j$ can provide $f$ with positive probability and the number of other players who can provide $f$ is weakly fewer than $b^*$.
    \end{lem}
    Lemma \ref{lem:hetupperbound} implies that $i$ can have at most $|F| b^*$ links.
\end{proof}
Proposition \ref{generalmprop} extends Theorem \ref{T1} to the case with a general favor provision matrix. Moreover, $B(\alpha,\underline p,G_f,\delta)$ does not depend on $M$ or $N$ and the cooperation in the society is bounded as long as the favors are substitutable. 


\subsection{The Set of Strongly Stable Networks with Transfers is Nonempty}
\label{sec:SSTnetworkexample}
\begin{lem}\label{lem:SSTnetworkexample}
    Suppose that $d_i(g) = B^*$ for all $i$. Then $g$ is SST.
\end{lem}
\begin{proof}
Suppose that $\{i,j\}$ violates SST of $g$ via $g'$ and $t =(t_i,t_j)$. As $d_i(g')\leq B^*+1$ and  $d_j(g')\leq B^*+1$, we will consider 3 cases.

 \textbf{\underline{Case 1: $d_i(g') \leq B^*$ and $d_j(g') \leq B^*$.}} As the only link in $g'$ that is not in $g$ is $ij$, both $i$ and $j$ and all their neighbors at $g'$ has at most $B^*$ links. If $t_i = t_j$, then $\tilde u_i(g',j,t_i,t_j) = u_i(g')$ and   $\tilde u_j(g',i,t_j,t_i) = u_j(g')$. Then, by Lemma \ref{lem:fullisbest}, $u_i(g) \geq u_i(g')$ and $u_j(g) \geq u_j(g')$, which means that this is not a violation of SST. If $t_i > t_j$, then  $\tilde u_i(g',j,t_i,t_j) < u_i(g')$ and thus $\tilde u_i(g',j,t_i,t_j) < u_i(g')$  which means that this is not a violation of SST. If $t_i < t_j$, then replacing $i$ and $j$ in the last sentence shows that this is not a violation of SST.

  \textbf{\underline{Case 2: $d_i(g') = B^*+1$ and $d_j(g') = B^*+1$.}} Without loss of generality, let $t_i \geq t_j$. 
  \begin{equation}
      \tilde u_i(g',j,t_i,t_j) - u_i(g'-ij) \leq u_i(g') - u_i(g'-ij)
  \end{equation}
From definition of $B^*$, $u_i(g') - u_i(g'-ij) < \frac{1-\delta}{\delta} (c-\gamma)$, which shows that link $ij$ is not sustainable at $g'$ and this is not a violation of SST.

    \textbf{\underline{Case 3: $d_i(g') = B^*+1$ and $d_j(g') \leq B^*$.}} First, note that $t_j > t_i$, as otherwise the link $ij$ would not be sustainable at $g'$. Let $\mathbb E_{g'} [\tilde t] = \mathbb E_{g'} [t_j] - \mathbb E_{g'} [t_i]$ and $n = d_j(g') \leq B^*$. For the link $ij$ to be sustainable at $g'$ for $i$, we need
    \begin{equation}\label{eq:prop4sustain}
        \alpha v (1-(1-p)^{B^*}) p - \alpha c \frac{1-(1-p)^n}{n} + \mathbb E_{g'} [\tilde t]  \geq \frac{1-\delta}{\delta} (c-\gamma)
    \end{equation}
    For $\tilde u_j(g',i,t_j,t_i) \geq u_j(g)$, the following inequality must be satisfied
    \begin{equation}
\begin{split}
            &\alpha v (1-(1-p)^n) - \alpha c \frac{1-(1-p)^{B^*}}{B^*} (n-1) - \alpha c \frac{1-(1-p)^{B^*+1}}{B^*+1} - \mathbb E_{g'} [\tilde t] \\
            &- \alpha v (1-(1-p)^{B^*}) +  \alpha c \frac{1-(1-p)^{B^*}}{B^*} B^* \geq 0
\end{split}
    \end{equation}
which is equivalent to
        \begin{equation}
\begin{split}
            &\alpha v (1-(1-p)^n) - \alpha c \frac{1-(1-p)^{B^*}}{B^*} n - \alpha c \frac{1-(1-p)^{B^*+1}}{B^*+1} - \mathbb E_{g'} [\tilde t]  \\
            &- \alpha v (1-(1-p)^{B^*}) +  \alpha c \frac{1-(1-p)^{B^*}}{B^*} (B^*+1) \geq 0
\end{split}
    \end{equation}

As for $n < B^*$, $\alpha v (1-(1-p)^n) - \alpha c \frac{1-(1-p)^{B^*}}{B^*} n$ is increasing in $n$, previous equation implies
        \begin{equation}
\begin{split}
            &\alpha v (1-(1-p)^{B^*}) - \alpha c \frac{1-(1-p)^{B^*}}{B^*} B^* - \alpha c \frac{1-(1-p)^{B^*+1}}{B^*+1} - \mathbb E_{g'} [\tilde t]  \\
            &- \alpha v (1-(1-p)^{B^*}) +  \alpha c \frac{1-(1-p)^{B^*}}{B^*} (B^*+1) \geq 0
\end{split}
    \end{equation}
    Cancelling terms and re-arranging
    \begin{equation}
        \alpha c \frac{1-(1-p)^{B^*}}{B^*} - \alpha c \frac{1-(1-p)^{B^*+1}}{B^*+1} \geq \mathbb E_{g'} [\tilde t] 
    \end{equation}
    Re-arranging Equation \ref{eq:prop4sustain} and plugging in $n = B^*$ to minimize RHS,
    \begin{equation}
         \mathbb E_{g'} [\tilde t]  \geq \frac{1-\delta}{\delta} (c-\gamma) - \alpha v (1-(1-p)^{B^*}) p + \alpha c \frac{1-(1-p)^{B^*}}{B^*}
    \end{equation}
    For both of these equation to hold, we need
    \begin{equation}
        \alpha v (1-(1-p)^{B^*}) p -  \alpha c \frac{1-(1-p)^{B^*+1}}{B^*+1} \geq \frac{1-\delta}{\delta} (c-\gamma)
    \end{equation}
    which contradicts the definition of $B^*$, thus no such $\tilde t$ exists.
    \end{proof}

\subsection{Heterogeneity in Values, Costs and Discount Factor}\label{morehet}
I now allow players to have different values for $v_i$, $c_i$ and $\delta_i$. Formally, player $i$ discounts future payoffs by $\delta_i$ and whenever players $i$ and $j$ are playing (where $i$ is player $1$ and $j$ is player $2$), they play the following game $G_f^{ij}$:

\begin{table}[H]
    \caption{Payoffs in the Favor Exchange Game $G_f^{ij}$}
\begin{center}
        \begin{tabular}{cc|c|c|}
      & \multicolumn{1}{c}{} & \multicolumn{1}{c}{$C$}  & \multicolumn{1}{c}{$D$} \\\cline{3-4}
      \multirow{2}*{}  & $A$ & $(v_i,-c_j)$ & $(0,-\gamma)$ \\\cline{3-4}
    \end{tabular}
\end{center}
    \label{table:FEgameHet}
\end{table}

I refer $(v_i,c_i,\delta_i)$ as the type of player $i$. For any $i$, let $B(\alpha,p,v_i,c_i,\delta_i)$ denote the cooperation bound in a society formed by players of type $(v_i,c_i,\delta_i) \in \mathcal V$, where $ \mathcal V$ is finite. We say that $i$'s type is lower than $j$'s type if $B(\alpha,p,v_i,c_i,\delta_i) < B(\alpha,p,v_j,c_j,\delta_j)$ and $i$'s type and $j$'s type are equivalent if $B(\alpha,p,v_i,c_i,\delta_i) = B(\alpha,p,v_j,c_j,\delta_j)$.\footnote{I also assume that if there exits $i$ such that $B(\alpha,p,v_i,c_i,\delta_i) = B$  there are at least $B$ other players with equivalent types.} Finally, I assume that there is a $\overline B$, $B(\alpha,p,v_i,c_i,\delta_i) \leq \overline B$ for some $\overline B$.

Theorem \ref{T2} can be extended to this setting.
    \begin{prop}\label{heterogeneity}
   For any $(\alpha,p,\{G_f^{ij}\}_{(i,j) \in N^2},\{\delta_i)\}_{i \in N})$, in any strongly stable networks, the fraction of players who attain their cooperation bound goes to 1 and the fraction of players who are linked with lower type players goes to $0$ as $N \to \infty$.
\end{prop}
\begin{proof}
The following lemma derives an upper bound on the number of players who do not attain their cooperation bound.
        
        \begin{lem}
        For any $k < \overline B$ there can be at most $k+1$ players with $k$ links who do not attain their cooperation bound.
        \end{lem}
        \begin{proof}
        Let $g$ denote a strongly stable network. For a contradiction, assume for some $k < \overline B$, there are $k'>k$ players with $k$ number of links who do not attain their cooperation bound. Then at least two of those players are not linked, let $i$ and $j$ denote those players. Then $g' = g \cup \{ij\}$  satisfies the conditions of Definition \ref{strongstability} and $g$ is not strongly stable.
        \end{proof}
        
        \begin{lem}
        If two players attain their cooperation bound and are linked with a player who have fewer links than them, then they must be linked.
        \end{lem}
        \begin{proof}
        Let $i$ and $j$ denote the two players who are linked with a player who have fewer links than them and who attain their cooperation bound. Let $k_i$ and $k_j$ denote these players ($k_i=k_j$ is possible). Then $g' = g \cup \{ij\} \setminus \{i k_i,j k_j\}$  satisfies the conditions of Definition \ref{strongstability} and $g$ is not stable.
        \end{proof}

 Note that if all players have types from $\mathcal V$, for any society with $N$ members, there can be at most $\sum_{b = 1}^{\overline B} (b+1)$ players who do not attain their cooperation bound, which does not depend $N$. Hence as $N \to \infty$, the fraction of players who attain their cooperation bounds (and fraction of players who have links with others who have fewer links than them) goes to $1$.

\end{proof}


Proposition \ref{heterogeneity} shows that the intuition in the homogeneous case generalizes to the case with heterogeneity and most of the players reach the cooperation bound of their own type. Moreover, only a bounded number of players can have relationships with lower type players, and thus the network is fully stratified as $N$ goes to $\infty$. This shows that the heterogeneity over the players may have important implications for the network structure and players payoffs.

\subsection{Legal Enforcement and Population Size}\label{sec:popsize}

In this appendix, $C(\gamma)$ represents the total cost of maintaining a legal system, and per player cost is given by $C(\gamma)/N$. Next proposition shows that legal enforcement is the optimal enforcement mechanism is large societies.

\begin{prop}\label{optimallegal:prop}
For each $N$, there exists  $\gamma^*(N)$ that maximizes the per period expected payoff under legal enforcement. There exists $\overline N$ and $\overline N(\kappa)$ such that
\begin{itemize}
    \item Legal enforcement gives players higher payoffs than pure bilateral enforcement if and only if $N \geq \overline{N}$.
    \item Legal enforcement gives players higher payoffs than community enforcement if and only if $N \geq \overline N(\kappa)$.
\end{itemize}
Moreover, if $\gamma^*(N)$ and $\gamma^*(N')$ are maximizers and $N > N'$, then $\gamma^*(N) \geq \gamma^*(N')$. 
\end{prop}

\begin{proof}
The utility under legal enforcement is given by
\begin{equation}
  U(N,\gamma) \equiv \alpha (v-c) \left(1-(1-p)^{B^*(\gamma)}\right) -\frac{C(\gamma)}{N}
\end{equation}

Since $B^*(\gamma)$ is increasing in $\gamma$ and is defined on integers and $C(\gamma)$ is continuous, $U(\cdot,\gamma)$ only has jump discontinuities, thus is discontinuous at most countably many points.

\begin{lem}
 $U(\cdot,\gamma)$ is upper-semi continuous in $\gamma$.
\end{lem}
\begin{proof}
From Corollary \ref{legalcharacterization:cor}, $B^*(\gamma)$ is increasing in $\gamma$. Since  $U(\cdot,\gamma)$ is increasing in $B^*(\gamma)$, at each point $U(\cdot,\gamma)$ is discontinuous, it jumps up. Since $U(\cdot,\gamma)$ is continuous at any other point, is is usc.
\end{proof}

Since $\lim_{\gamma \to c} C(\gamma) = \infty$, and the first term of $  U(N,\gamma)$ is bounded, there exists $\epsilon>0$, such that $U(N,\gamma) < U(N,c-\epsilon)$ for all $\gamma > c-\epsilon$. Then the existence of an optimal $\gamma^*(N)$ follows from the compactness of $[0,c-\epsilon]$ and upper semi-continuity of $U(\cdot,\gamma)$.

To prove the second part, for a contradiction, assume that $N > N'$ but $\gamma^*(N) < \gamma^*(N')$, where $\gamma^*(N)$ and  $\gamma^*(N')$ are maximizers of $U(N,\gamma)$ and $U(N',\gamma)$. From optimality of $\gamma^*(N')$, we have $U(N',\gamma^*(N')) -  U(N',\gamma^*(N)) \geq 0$. Moreover
\begin{equation}
\begin{split}
      &\left(U(N,\gamma^*(N')) -  U(N,\gamma^*(N))\right) - \left(U(N',\gamma^*(N')) -  U(N',\gamma^*(N))\right)\\
      =&- \left( \frac{C(\gamma^*(N'))}{N} - \frac{C(\gamma^*(N))}{N} \right) + \left( \frac{C(\gamma^*(N'))}{N'} - \frac{C(\gamma^*(N))}{N'} \right) > 0\\
      > 0
\end{split}
\end{equation}

where the inequality follows from $N' < N$ and $C(\gamma^*(N')) - C(\gamma^*(N)) > 0$ and implies $\left(U(N,\gamma^*(N')) -  U(N,\gamma^*(N))\right) > 0$, contradicting the optimality of $\gamma^*(N)$.

To compare the two bilateral enforcement mechanisms, observe that $\max_{\gamma} U(\gamma,N) =  U(\gamma^*(N),N)$ is increasing in $N$ since $ U(\gamma^*(N),N) \geq U(\gamma^*(N'),N) >U(\gamma^*(N'),N')  $ when $N > N'$. The result then follows from observing the optimality of legal enforcement when $N \to \infty$ and bilateral enforcement when $N \leq B^*(0)$.

To compare the efficiency of legal enforcement  and community enforcement, I first show that as $N \to \infty$, legal enforcement can approximate the first best payoff of $\alpha (v-c)$ arbitrarily.
\begin{lem}
$\lim_{N \to \infty} \max_{\gamma} U(N,\gamma) = \alpha (v-c)$
\end{lem}
\begin{proof}
Let $\epsilon$ be given. First, note that there exists $\hat \gamma$ such that $\alpha(v-c)(1-(1-p)^{\hat \gamma}) > \alpha (v-c) - \epsilon/2$. Next, observe that there exists $\hat N(\hat \gamma)$ such that for all $N \geq \hat N$, $C(\hat \gamma)/N < \epsilon/2$, which proves the result.
\end{proof}

Let $U(B(\kappa),\kappa)$ denote payoff at the optimal community size under $\kappa$. Clearly, $U(B(\kappa),\kappa) <  \alpha (v-c)$. The result than follows from the previous lemma and the fact that  $\max_{\gamma} U(\gamma,N) =  U(\gamma^*(N),N)$ is increasing in $N$.
\end{proof}

\end{appendices}

\end{document}